\documentclass[11pt, usenames,dvipsnames]{article}
\pdfoutput=1

\usepackage{bm,wrapfig,float,array,mathtools}
\usepackage{amsfonts}
\usepackage{amssymb}
\usepackage{cite}
\usepackage{amsmath}
\usepackage{color}
\usepackage{graphicx}
\usepackage{hyperref}
\usepackage{float}
\usepackage[T1]{fontenc}
\usepackage[utf8]{inputenc}
\usepackage{alphabeta}
\usepackage{fancyvrb}
\usepackage[dvipsnames]{xcolor}
\usepackage{bold-extra}
\usepackage{lmodern}
\usepackage{cleveref}
\usepackage[normalem]{ulem}
\usepackage{tikz-cd}
\newcommand{\RNum}[1]{\uppercase\expandafter{\romannumeral #1\relax}}
\usepackage[dvipsnames]{xcolor}
\usepackage[framemethod=TikZ]{mdframed}
\usepackage{amsthm}

\usepackage{multirow}

\usepackage{subfigure}
\usepackage{paralist}
\usepackage{tikz}
\usepackage{diagbox}
\usetikzlibrary{positioning}
\usetikzlibrary{calc}
\usetikzlibrary{decorations.pathreplacing,calligraphy}

\usepackage{xstring}
\usetikzlibrary{decorations.pathmorphing} 
\usetikzlibrary{decorations.markings} 
\usetikzlibrary{arrows} 
\usetikzlibrary{shapes} 
\usetikzlibrary{matrix} 
\usetikzlibrary{positioning} 
\usepackage[english]{babel} 
\usepackage[autostyle]{csquotes}
\usetikzlibrary{shapes.multipart}
\usetikzlibrary{patterns}

\newcommand{\dd}{\mathop{}\!\mathrm{d}}

\tikzstyle{brane}=[draw]
\tikzset{D7/.style={circle, draw=black, inner sep=0pt, fill=white, minimum size=3mm}}
\tikzset{hasse/.style={circle, fill,inner sep=2pt}}
\tikzset{flavor/.style={regular polygon,fill=white,regular polygon sides=4,inner sep=2.5pt, draw}}
\tikzset{gauge/.style={circle, draw,inner sep=2.5pt}}
\tikzset{gaugeb/.style={circle, draw,fill=black,inner sep=2.5pt}}
\tikzset{gauger/.style={circle, draw,fill=cyan,inner sep=2.5pt}}
\tikzset{gaugeg/.style={circle, draw,fill=red,inner sep=2.5pt}}
\tikzset{SUd/.style={circle, draw=black, inner sep=0pt, fill=yellow, minimum size=2mm}}
\tikzset{bd/.style={circle, draw=black, inner sep=0pt, fill=black, minimum size=2mm}}
\tikzset{wd/.style={circle, draw=black, inner sep=0pt, fill=white, minimum size=2mm}}
\tikzset{Dynkin/.style={circle, draw=black, inner sep=0pt, fill=white, minimum size=2mm}}
\tikzstyle{ligne}=[draw, thick] 
\tikzset{doublearrow/.style={ draw=black!75, color=black!75, thick, double distance=3pt, }}

\numberwithin{equation}{section}  

\graphicspath{ {figs/} }

\newcommand{\be}{\begin{equation}}
\newcommand{\ee}{\end{equation}}
\newcommand{\ba}{\begin{aligned}}
\newcommand{\ea}{\end{aligned}}

\newcommand{\bra}[1]{\langle #1 |}
\newcommand{\ket}[1]{| #1 \rangle}
\newcommand{\braket}[2]{\langle #1 | #2 \rangle}

\begin{document}

\baselineskip=18pt  
\numberwithin{equation}{section}  
\allowdisplaybreaks  


%
%


\thispagestyle{empty}

\vspace*{0.8cm} 
\begin{center}
{\huge SymTh for non-finite symmetries }\\

 \vspace*{1.5cm}
{\large Fabio Apruzzi$^{1}$, Francesco Bedogna$^{1}$, Nicola Dondi$^{2}$} \
 \vspace*{.2cm}
 \smallskip

{\it $^1$ Dipartimento di Fisica e Astronomia “Galileo Galilei”, Università di Padova,\\ Via Marzolo 8, 35131 Padova, Italy  }\\

\smallskip

{\it $^1$ INFN, Sezione di Padova Via Marzolo 8, 35131 Padova, Italy   }\\

\smallskip

{\it $^2$Abdus Salam International Centre for Theoretical Physics, \\
Strada Costiera 11, 34151, Trieste, Italy.   }\\
{\it $^2$ INFN, Sezione di Trieste, Via Valerio 2, I-34127 Trieste, Italy.  }\\

\vspace*{2cm}
\end{center}

\noindent
Symmetry topological field theory (SymTFT) is a convenient tool for studying finite generalized symmetries of a given quantum field theory (QFT). In particular, SymTFTs encode all the symmetry structures and properties, including anomalies. Recently, this tool has been applied for non-finite symmetries as well. In this paper, we take a different route, which consists of considering a free theory rather than a topological field theory in the bulk. We call it Symmetry Theory (SymTh). We study its topological operators together with the free boundary conditions. We also propose a procedure that is analogous to the sandwich construction of SymTFTs and allows us to obtain the physical QFT. We apply this to many examples, ranging from abelian $p$-form symmetries to 2-groups, and the (solvable) case of group-like symmetries in quantum mechanics. Finally, we provide a derivation of the SymTh of $\mathbb Q/ \mathbb Z$ non-invertible symmetries from the dimensional reduction of IIB supergravity on the conifold. In addition, we give an ultraviolet interpretation of the quantum Hall states dressing the non-invertible $\mathbb Q/ \mathbb Z$ topological defects, in terms of branes in the IIB supergravity background.

 \newpage

\tableofcontents


\newpage

\section{Introduction}
Symmetry Topological Field Theories (SymTFTs) play a crucial role in the study of finite generalized symmetries in quantum field theories (QFTs) \cite{Gaiotto:2014kfa, Gaiotto:2020iye, Apruzzi:2021nmk, Freed:2022qnc, Apruzzi:2022rei, Kaidi:2022cpf, Antinucci:2022vyk, Bhardwaj:2023ayw, Kaidi:2023maf, Zhang:2023wlu, Bhardwaj:2023fca, Bartsch:2023wvv, Apruzzi:2022dlm, Schafer-Nameki:2023jdn}. In particular, they capture the symmetry data of a given \(d\)-dimensional QFT. A SymTFT consists of a topological theory in a \((d+1)\)-dimensional bulk, accompanied by gapped boundary conditions and, potentially, a physical boundary. When the SymTFT is defined on an interval, there are two boundaries. A physical QFT is defined at one boundary while the other remains gapped. Upon interval compactification, the two boundaries merge, and this procedure recovers the \(d\)-dimensional QFT under consideration. The topological operators of the SymTFT can be inserted to describe the symmetries of the \(d\)-dimensional QFT, their actions on charged operators and various twisted sectors. In summary, the SymTFT serves as a valuable framework that encodes all the finite symmetry properties of the QFT in question.

Recently, there have been proposals for generalizations of the SymTFT to continuous global symmetries \cite{Brennan:2024fgj, Antinucci:2024zjp, Bonetti:2024cjk}. For abelian symmetries, these generalizations consist of BF-theories where the gauge fields are valued in \( U(1) \) and \( \mathbb{R} \), rather than in a finite group. The SymTFT for the non-abelian case is slightly more intricate, involving a BF theory originally proposed in \cite{Horowitz:1989ng}. This theory includes a compact gauge field and as  many \( \mathbb{R} \) fields as the dimension of the Lie algebra. For one-dimensional QFTs and a two-dimensional SymTFT, the topological operators of the bulk theory can be interpreted in terms of a free non-abelian Yang-Mills (YM) theory in the bulk. The equivalence between the TQFT proposed in \cite{Horowitz:1989ng} and the two-dimensional YM theory was established in \cite{Witten:1992xu}. A generalization of this framework to higher dimensions has been recently proposed in \cite{Bonetti:2024cjk}.

In this paper, we advocate an alternative approach to describing the (generalized) symmetries of a given QFT. This approach involves a bulk Symmetry Theory, which we refer to as SymTh, that is not topological. This framework is inspired by holographic or geometric constructions in string theory. In such cases, the bulk description is typically provided by a collection of weakly coupled Maxwell theories, coupled via Chern-Simons interactions \cite{Hofman:2017vwr, Damia:2022seq, Damia:2022bcd}. Unlike topological theories, the bulk theory in this context depends on a coupling constant, is not topological, and is subject to renormalization. We focus on an effective regime where the bulk theory is weakly coupled. The advantage of this proposal is that it naturally captures non-flat configurations of continuous symmetries. These configurations are not automatically described by the SymTFT, which encodes flat configurations. However, \cite{Brennan:2024fgj, Antinucci:2024zjp} have proposed alternative methods to incorporate non-flat configurations within the SymTFT framework.

The goal of this paper is to extract the (non-finite) symmetry sector of a given QFT. To achieve this, we focus on studying the topological operators of the SymTh, along with the free (non-gapped) boundary conditions. We also explore the interplay between boundary conditions and the parallel projections of the topological operators onto the boundary. An additional advantage of this approach is the ability to consider modified Neumann boundary conditions, which facilitate the dynamical gauging of continuous symmetries. This flexibility is one of the key strengths of our framework. Finally, we propose a procedure to define the interval compactification, thereby providing an analogy to the sandwich construction of the SymTFT. This is accomplished by carefully factorizing the bulk divergent part of the partition function, which can be interpreted as decoupling the bulk physics.

We examine several examples, beginning with the model that describes a \( p \)-form \( U(1) \) symmetry, which corresponds to a \( (p+1) \)-form Maxwell theory in the bulk. We carefully review its boundary conditions and topological operators. For this example, we explicitly construct the sandwich procedure, which facilitates interval compactification and effectively decouples the bulk from the boundary QFT. Next, we explore lower-dimensional examples, focusing on as quantum mechanics (QM) with abelian symmetry and mentioning extensions to non-abelian symmetries. In this case, we propose 2d Maxwell and 2d Yang-Mills theories as SymTh and investigate their properties. Since we can solve the 2d theory, it is possible to compute the path integral over bulk fields exactly, demonstrating that it reproduces various symmetry properties of the boundary QM, depending on the operators inserted in the bulk. We also discuss two different examples of 2-groups and propose the SymTh for 0- and 1-form \( \mathbb{Q}/\mathbb{Z} \) symmetries \cite{Choi:2022fgx, Choi:2022jqy, Damia:2022bcd, Damia:2022seq, Cordova:2022ieu, Putrov:2022pua, Shao:2023gho}. In this setup, we show how to derive the Chern-Simons terms from the 4d axion-Maxwell Lagrangian, coupled to background fields. Additionally, we provide a derivation of the SymTh by reducing IIB supergravity on the boundary of the conifold, \( T^{1,1} \cong S^2 \times S^3 \). Finally, we discuss how to use the no global symmetry hypothesis in quantum gravity to argue, in general, for the presence of branes that generate finite symmetries. This hypothesis can be employed to identify the branes responsible for the quantum Hall state, which dresses topological non-invertible operators. This is indeed the SymTh of the 4d axion-Maxwell theory.

The paper is organized as follows. In Section \ref{sec:maxw}, we propose the \( (p+1) \)-form Maxwell theory as the symmetry theory for a \( U(1)^{(p)} \) symmetry and study its properties in detail. In Section \ref{eq:ldex}, we examine lower-dimensional examples, focusing on the SymTh of a (non)-abelian symmetry in a generic quantum mechanics system. In Section \ref{sec:hdex}, we consider higher-dimensional examples, with a focus on 2-groups. In Section \ref{sec:bulkaxmax}, we provide a derivation for the SymTh of 0- and 1-form non-invertible symmetries and study their properties. Finally, in Section \ref{sec:branes}, we offer a bottom-up perspective on how branes can provide a UV avatar for topological defects of discrete symmetries, as well as quantum Hall states dressing topological operators.

\paragraph{Note added:} 
During the preparation of this work, two preprints \cite{Brennan:2024fgj, Antinucci:2024zjp} appeared. The intent of these papers is very similar to ours, though the methodology employed differs. During the completion of this work, \cite{Bonetti:2024cjk} appeared, which contains a similar methodology for the case of non-abelian 0-form symmetries.

\section{WarmUp: SymTh for a $U(1)$ $p$-form global symmetry}
\label{sec:maxw}

In this section, we propose a symmetry theory for a \( U(1) \) \( p \)-form global symmetry of a QFT at the boundary. To do so, we first review some known aspects that are ubiquitous in holographic and string-theoretic constructions. In particular, in the case of holography, the symmetry theory corresponds to the truncation of the full bulk theory to the sector that describes only the behavior of the symmetry, possibly with contributions from localized brane singularities. For string theory constructions that describe the symmetry sector, we instead focus on the flux sector and its reduction on the link \( L \), eventually decorated by fields corresponding to isometries of \( L \), with the possibility of contributions from the resolution of singularities on \( L \). This procedure was highlighted in \cite{Apruzzi:2023uma}.

Differently from the symmetry TFT construction of discrete symmetries, the proposal of this paper does not consist of a symmetry topological field theory in the bulk. In particular, we claim that the bulk kinetic terms play an important role in the discussion. Nevertheless, even though the theory is not topological, we will focus on its topological properties, as these are key to capturing the symmetries of the boundary QFT. Therefore, even if the bulk theory is not preserved under RG-flow and can be considered an effective bulk description, its topological aspects are still preserved and universal. This strategy has been adopted many times to describe symmetries from a bulk perspective, and more recently in \cite{Damia:2022seq, Damia:2022bcd}.

The description that we propose here for the SymTh of a \( U(1) \) \( p \)-form global symmetry consists of a QFT in \( d \) dimensions and a \( (p+1) \)-form Maxwell theory in \( (d+1) \) dimensions:
\begin{equation} \label{eq:Max}
    S_{d+1} = -\frac{1}{2g^2} \int_{M_{d+1}} da_{p+1} \wedge \ast da_{p+1} \, ,
\end{equation}
where we have reabsorbed the coupling for convenience. We will reinstate the explicit dependence on the coupling when needed in the following. Again, we are primarily interested in the topological properties of this action. We will now study the topological operators of this theory and how they define the symmetries of the \( d \)-dimensional boundary QFT, depending on the boundary conditions for the \( a_{p+1} \) field. In the context of holography and string theory, we have a natural candidate for the bulk direction, sometimes called the radial direction. The space where the QFT lives is given by \( M_d = \partial M_{d+1} \). In holography, we usually work in Anti-de-Sitter (AdS) space \cite{Hofman:2017vwr, Damia:2022seq, Damia:2022bcd}, and we have a single boundary where we impose boundary conditions. The physical theory at the boundary is dual to the full bulk gravity. As explained in \cite{Heckman:2024oot, Apruzzi:2023uma}, one can think of the physical boundary as being smeared in the bulk, so there is no physical separation between the physical boundary and the boundary condition for the fields. To resolve the physical boundary from the topological boundary, we truncate the full bulk theory to the Maxwell sector and its Chern-Simons coupling. We then consider the space \( M_{d+1} \) to be flat. Finally, we attempt to generalize the sandwich picture of the SymTFT to capture the properties of the SymTh, highlighting the key differences. To do so, we need \( M_{d+1} = M_d \times I \), where \( I = [0,L] \) is a finite interval, and there are two boundaries \( M_d \) at \( x = 0 \) and \( x = L \), if we denote \( x \) as the coordinate of the interval.

\subsection{Topological operators and their action}
\label{sec:setup}
\begin{figure}
\centering
\begin{tikzpicture}
\begin{scope}[shift={(0,0)}]
\draw [orange, thick, fill=yellow,opacity=0.3] 
(0,0) -- (0, 4) -- (2, 5) -- (2,1) -- (0,0);
\draw [yellow, thick]
(0,0) -- (0, 4) -- (2, 5) -- (2,1) -- (0,0);
\node at (1,3.5) {$\mathcal B$};
\end{scope}
\draw[dashed] (0,0) -- (4,0);
\draw[dashed] (0,4) -- (4,4);
\draw[dashed] (2,5) -- (6,5);
\draw[dashed] (2,1) -- (6,1);
\draw [Red] (4,3) ellipse (0.2 and 0.8) ;
    \node [Red] at (5.3,2.5) {$U_{\alpha}(\Sigma_{d- p -1})$};
\draw[thick, Green] (1,3) -- (5,3) ;
  \node [Green] at (1,2.6) {$\mathcal E(M_p)$};
\draw [Green,fill=Green] (1,3) ellipse (0.05 and 0.05);
 \node at (2.9,4.5) {SymTh};
  \node[Green] at (2.9,3.3) {$W_q(M_{p+1})$};
\draw [Purple] (4,1) ellipse (0.2 and 0.8) ;
  \node [Purple] at (5.1,1.4) {$U_{\beta}(\Sigma_{p+2})$};
\draw[thick, Blue] (3.8,1) ellipse (0.3 and 0.2) ;
  \node[Blue] at (2.6,0.5) {$V_m(M_{d-p-2})$};
\end{tikzpicture}
\caption{Topological operators $U$ linking with Wilson surfaces $W$ or $V$. Depending on boundary conditions the Wilson surface can end on the boundary or not. If the Wilson surface $W$ (for example) ends on the boundary it will source a non-topological operator, $\mathcal E$. In this case, the topological operator $U$ defines a symmetry of the boundary theory. 
\label{fig:ggg}}
\end{figure}
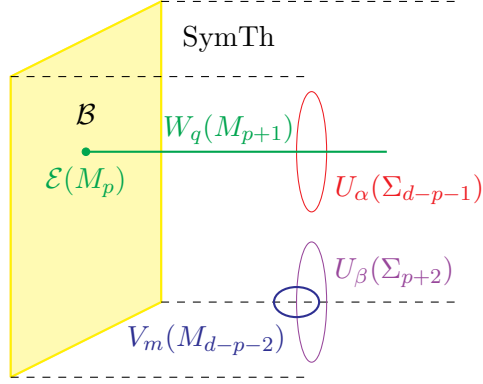

We will review in detail the construction of topological operators in the theory \eqref{eq:Max} to set the stage for what follows. There are two conserved currents:
\begin{equation}
    d \ast_{d+1} J_{p+2} = -\frac{1}{g^2} d \ast_{d+1} f_{p+2} = 0 \qquad  d \ast J_{d-p-1} = \frac{1}{2\pi }d f_{p+2}= 0 \, ,
\end{equation}
where \( f_{p+2} = da_{p+1} \), and the conservation comes from the Bianchi identity and the equation of motion, respectively. The currents are given by 
\begin{equation}
  J_{p+2} = -\frac{f_{p+2}}{g^2}, \qquad   J_{d-p-1} = \frac{i}{2\pi } (-)^{(d+1)(p-d+1)} \ast_{d+1} f_{p+2} \, .
\end{equation}
The topological operators are defined as follows:
\begin{equation} \label{eq:topopmax}
\begin{aligned}
  &  U_{\alpha}(\Sigma_{d- p -1}) = e^{ i \alpha \int_{\Sigma_{d- p -1}} \ast_{d+1} J_{p+2} } = e^{ -i \alpha \int_{\Sigma_{d- p -1}} \ast_{d+1} \frac{f_{p+2}}{g^2}} \\ 
  & U_{\beta}(\Sigma_{p+2}) = e^{ i \beta \int_{\Sigma_{p+2}} \ast_{d+1} J_{d-p-1}} = e^{ i \beta \int_{\Sigma_{p+2}} \frac{f_{p+2}}{2\pi}} \,,
\end{aligned}
\end{equation}
where \( \alpha \) and \( \beta \) are defined between \( [0, 2 \pi) \). The current conservation ensures that these operators are topological. These bulk topological operators charge the Wilson surface operators, which are defined by
\begin{equation} \label{eq:WSmax}
\begin{aligned}
    & W_q(M_{p+1}) = e^{ i q \int_{M_{p+1}} a_{p+1}} \\
    & V_m(M_{d-p-2}) = e^{ i m \int_{M_{d-p-2}} b_{d-p-2}} \, ,
\end{aligned}
\end{equation}
where \( b_{d-p-2} \) is the field magnetic dual\footnote{We can in principle dualize the bulk theory by using a Lagrange multiplier field \( b_{d-p-2} \) that induces the Bianchi identity when integrated out,
\begin{equation}
  S_{d+1}^{\rm dual}=  -\int_{M_{d+1}} \frac{1}{2g^2}f_{p+1}\wedge \ast f_{p+1}- f_{p+1}\wedge db_{d-p-2}.
\end{equation}
If we integrate out \( f_{p+1} \), we get that \( \ast f_{p+1} = db_{d-p-2} \). The action reads \( S_{d+1}^{\rm dual}=  -\frac{1}{2}\int_{M_{d+1}}db_{d-p-2}\wedge \ast db_{d-p-2} \). In addition, it is possible to rewrite all the topological operators in terms of the dual field \( b_{d-p-2} \).}  \( a_{p+1} \). Differently from the finite case, which is, for instance, described by a BF theory, the Wilson surface operators are not topological. The topological operators act on the Wilson surfaces by linking as follows:
\begin{equation} \label{eq:action_on_Wsurf}
\begin{aligned}
   & \langle  U_{\alpha}(\Sigma_{d- p -1}) W_q(M_{p+1}) \rangle = e^{  i q \alpha \text{Link}(\Sigma_{d- p -1},M_{p+1}) } \langle  W_q(M_{p+1}) U_{\alpha}(\widetilde{\Sigma}_{d- p -1})  \rangle \\
 &   \langle  U_{\beta}(\Sigma_{p+2}) V_m(M_{d - p-2}) \rangle = e^{  i m  \beta \text{Link}(\Sigma_{p+2},M_{d - p-2}) } \langle  V_m(M_{d - p-2}) U_{\beta}(\widetilde{\Sigma}_{p+2}) \rangle \, ,
\end{aligned}
\end{equation}
where \( \widetilde{\Sigma} \) is the deformed version of \( \Sigma \). The boundary condition will determine which Wilson surfaces can extend radially and end at the boundary. We will see that this determines the topological surface that gives a genuine symmetry of the boundary QFT, see Figure~\ref{fig:ggg}, and which one does not act faithfully.

\subsection{Boundary conditions}

The set of boundary conditions for the free \((d+1)\)-dimensional \((p+1)\)-form Maxwell theory is reviewed in \cite{Lake:2018dqm}. This set comprises all boundary conditions such that the boundary variation of the free \((p+1)\)-form Maxwell action vanishes:
\begin{equation} \label{eq:bS}
    \delta S|_{\partial M_{d+1}} =- \frac{1}{g^2}\int_{\partial M_{d+1}} \delta a_{p+1} \wedge \ast_{d+1} f_{p+2} = 0 \,.
\end{equation}
This is solved by the Dirichlet boundary condition:
\begin{equation}
   D(a_{p+1}):\quad  \delta a_{p+1}|_{\partial M_{d+1}} = 0 \quad \Leftrightarrow \quad h_{y i_1 \ldots i_{d-p-2}}|_{\partial M_{d+1}} = 0 \, ,
\end{equation}
where the dual field is \( h_{d-p-1} = db_{d-p-2} = -\frac{1}{g^2}\ast_{d+1} f_{p+2} \). This implies that the gauge transformation \( a_{p+1} \rightarrow a_{p+1} + d\lambda_p \) vanishes at the boundary, allowing the Wilson surfaces \( W_q(M_{p+1}) \) to end on it. The second expression is the boundary condition expressed in coordinates, as in \cite{DiPietro:2019hqe}, where \( y \) is the radial coordinate and \( i_1, \ldots, i_d \) label the coordinates on \( M_d \). If we denote \( \mathcal{B} \) as a connected component of \( \partial M_{d+1} \) where we impose Dirichlet conditions, it means that the parallel (or complete) projection of \( U_{\alpha}(\Sigma_{d-p-1}) \) on \( \mathcal{B} \) generates the \( U(1)^{(p)} \) symmetry operator:
\begin{equation}
    {\rm Proj}(U_{\alpha}(\Sigma_{d-p-1})) \in U(1)^{(p)} \, ,
\end{equation}
whereas, since \( V_m(M_{d-p-2}) \) cannot end on the boundary, the parallel projection of \( U_{\beta}(\Sigma_{p+2}) \) trivializes and acts as the identity operator on the free boundary \( \mathcal{B} \), as shown in figure \ref{fig:ggg}. In general, the parallel projection does not give a simple object at the boundary, but in the case discussed in this paper, the projection leads to simple objects.

Alternatively, \eqref{eq:bS} is solved by the Neumann boundary condition, where \( a_{p+1} \) is allowed to vary freely at the boundary, subject to the constraint that
\begin{equation}
    N(a_{p+1}):\quad-\frac{1}{g^2}\ast_{d+1} f_{p+2}|_{\partial M_{d+1}} = 0 \quad \Leftrightarrow \quad f_{y i_1 \ldots i_{p+1}}|_{\partial M_{d+1}} = 0 \, .
\end{equation}

In the dual frame, we would instead have
\begin{equation}
    \delta S^{\rm dual} = -\frac{1}{\tilde g^2}\int_{\partial M_{d+1}} \delta b_{d-p-2} \wedge \ast_{d+1} h_{d-p-1} \, ,
\end{equation}
where $\tilde g^2 = \frac{1}{g^2}$.
The condition \( \delta S^{\rm dual} = 0 \) is solved by
\begin{equation}
   D(b_{d-p-2}):\quad  \delta b_{d-p-2}|_{\partial M_{d+1}} = 0 \quad \Leftrightarrow \quad f_{y i_1 \ldots i_{p+1}}|_{\partial M_{d+1}} = 0 \, .
\end{equation}
This implies that the gauge transformation \( b_{d-p-2} \rightarrow b_{d-p-2} + d\lambda_{d-p-3} \) vanishes at the boundary, allowing the Wilson surfaces \( V_m(M_{d-p-2}) \) to end on it. This means that the parallel (or complete) projection of \( U_{\beta}(\Sigma_{p+2}) \) on the boundary \( \mathcal{B} \) with Dirichlet conditions generates the \( U(1)^{(d-p-3)} \) symmetry:\footnote{We remark that this is the symmetry that is electromagnetic dual to the $p$-form symmetry in the bulk and not in the boundary.}
\begin{equation}
    {\rm Proj}(U_{\beta}(\Sigma_{p+2})) \in U(1)^{(d-p-3)} \, ,
\end{equation}
whereas, since \( W_q(M_{d+1}) \) cannot end on the boundary, the parallel projection of \( U_{\alpha}(\Sigma_{d-p-1}) \) trivializes acting as the identiry operator on the free boundary \( \mathcal{B} \), as shown in figure \ref{fig:ggg}.

Alternatively, the boundary problem is solved by the Neumann boundary condition, where \( b_{d-p-2} \) is allowed to vary freely at the boundary, subject to the constraint that
\begin{equation}
    N(b_{d-p-2}):\quad -\frac{1}{\tilde g^2}\ast_{d+1} h_{d-p-1}|_{\partial M_{d+1}} = 0 \quad \Leftrightarrow \quad h_{y i_1 \ldots i_{d-p-2}}|_{\partial M_{d+1}} = 0 \, .
\end{equation}
We can observe from the coordinate expression that
\begin{equation}
     D(a_{p+1}) \equiv N(b_{d-p-2}), \qquad N(a_{p+1}) \equiv D(b_{d-p-2}) \, ,
\end{equation}
namely, Neumann and Dirichlet conditions are dual under the Hodge star. In addition, \( D(a_{p+1}) \) and \( N(a_{p+1}) \) are mutually exclusive since Dirichlet imposes a fixed value at the boundary for the \( U(1) \) gauge field \( a_{p+1} \), whereas \( N(a_{p+1}) \) allows \( a_{p+1} \) to vary freely. Both conditions cannot be satisfied simultaneously on the same boundary. Finally, these boundary conditions are all free (or gapless) rather than gapped since the bulk fields will induce a free theory living on the boundary, which is given by the free dual gauge field \( b_{d-p-2} \) or the gauge field \( a_{p+1} \) for Dirichlet and Neumann boundary conditions, respectively.

We can also have a boundary that is not a free theory but has some non-trivial dynamics and interactions, i.e., a generic QFT in \(d\) dimensions with a \(U(1)\) \(p\)-form global symmetry. The coupling with the bulk field \(a_{p+1}\) is specified by the boundary term \cite{DiPietro:2019hqe},
\begin{equation}\label{eq:J-a_coupling}
    \int_{M_d} a_{p+1} \wedge \ast_d J^{QFT}_{p+1} + \ldots \, ,
\end{equation}
where \(\ldots\) can be additional couplings such as seagull terms. Suppose that \(M_{d+1} = M_d \times I\), where \(I = [0, L]\). There are two boundaries, and we place the physical one at \(x = 0\). The total action reads
\begin{equation}
    S_{\rm tot} = -\frac{1}{2g^2} \int_{0 < x < L} \int_{M_d} da_{p+1} \wedge \ast_{d+1} da_{p+1} + \int_{M_d(x=0)} a_{p+1} \wedge \ast_d J^{\rm QFT}_{p+1} + \ldots \, ,
\end{equation}
where \(\ldots\) represents boundary terms that do not depend on \(a_{p+1}\). By varying with respect to \(a_{p+1}\), we get
\begin{equation}
\begin{aligned}
    \delta S_{\rm tot} = & -\frac{1}{2g^2} \int_{0 < x < L} \int_{M_d} \delta a_{p+1} \wedge d \ast_{d+1} da_{p+1} \\
    & + \int_{M_d(x=0)} \delta a_{p+1} \wedge \ast_d J^{\rm QFT}_{p+1} \\
    & +\frac{1}{2g^2}  \int_{M_d(x=L)} \left(\delta a_{p+1} \wedge \ast_{d+1} f_{p+2}\right)|_{x=L} -\frac{1}{2g^2} \int_{M_d(x=0)} \left(\delta a_{p+1} \wedge \ast_{d+1} f_{p+2}\right)|_{x=0} \, .
\end{aligned}
\end{equation}
The first term gives the bulk equation of motion \(d \ast_{d+1} da_{p+1} = 0\). The second and last terms impose the modified Neumann boundary condition at \(x = 0\),
\begin{equation}
    \label{eq:constMNeumann}
    \ast_d J^{QFT}_{p+1} = \ast_{d+1} J_{p+2} |_{\partial M_{d+1}} = -\frac{1}{g^2} \ast_{d+1} f_{p+2} |_{M_d(x=0)} \, .
\end{equation}
The third term will give the boundary conditions at \(x = L\), where we have the choice to impose either the standard Neumann or Dirichlet boundary conditions. In the next subsection, we will study what happens when we take the \(L \to 0\) limit and how the two boundaries interact. In particular, we will give a prescription for a sandwich picture \cite{Gaiotto:2020iye, Kaidi:2022cpf} similar to the one for SymTFTs, by highlighting the key differences of having a gapless theory in the bulk rather than a topological theory. The goal is to isolate the symmetry sector of the physical QFT at the boundary. Therefore, while we will be interested in the free boundary conditions, we will try to factor out the dynamics of the gauge field in the bulk. Very importantly, we will also focus on the topological operators of the bulk theory and how they realize the symmetries of the physical boundary QFT.

\subsubsection*{Non-genuine defects} In the case of Neumann boundary conditions for \(a_{p+1}\), we cannot terminate its Wilson surface \(W_q(M_{p+1})\) on the boundary due to gauge invariance. However, the explicit Neumann boundary condition
\begin{equation}
    \partial_x a_{i_1,\ldots,i_{p+1}} = 0
\end{equation}
allows for L-shaped configurations \cite{Apruzzi:2023uma}. This leads to non-genuine \(\mathcal{O}(M_{p})\) operators , living at the end of \(W_q(M_{p+1})\). 

When we impose Neumann boundary conditions for \(b_{d-p-2}\), we cannot terminate the Wilson surface \(V_m(M_{d-p-2})\) on the boundary due to gauge invariance. However, the explicit Neumann boundary condition
\begin{equation}
    \partial_x b_{i_1,\ldots,i_{d-p-2}} = 0
\end{equation}
also allows for L-shaped configurations. This leads to the non-genuine \(\mathcal{O}(M_{d-p-3})\) operators, living at the end of \(V_m(M_{d-p-2})\). 

We recall that the charged operators $W_q, V_m$ in the SymTh are not topological themselves. Their L-shaped configuration leads to non-genuine operators attached to non-topological higher-dimensional defects. So far we have focused on this because they have a direct relation with the symmetry operators via linking. One can also consider L-shaped configuration formed by the topological defects  $U_{\alpha}$ or $U_{\beta}$. Depending on the boundary condition either $U_{\alpha}$ or $U_{\beta}$ trivializes at the boundary. The other one can also form an L-shaped configuration leading to non-genuine operators which attaches to a topological surfaces. It would be interesting to further study these ones in the future.  

\subsubsection*{Gauging and Neumann}
In the case of \(a_{p+1}\) having a Neumann boundary condition, the explicit gauging of the \(U(1)^{(p)}\) is not manifest. To understand the theory living on the free Neumann boundary, we need to look at the bulk description and implement an expansion close to the Neumann boundary. Let us call the expansion parameter \(\ell\), i.e., a small interval close to the Neumann boundary. The expansion depends on the bulk metric. In this case, we work with a flat metric, but it will change in the case of Anti-de-Sitter \cite{Hofman:2017vwr, Damia:2022bcd, Damia:2022seq}. By imposing Neumann boundary conditions, we get
\begin{equation}
    S_{\rm bulk} = -\frac{\ell}{2 g^2} \int_{M_d} da_{p+1} \wedge \ast_d da_{p+1} + O(\ell^2) + \ldots
\end{equation}
This is the limit where shrinking the interval direction will lead to a strongly coupled kinetic term. There is a way to avoid this issue that is related to the singleton sector discussed in \cite{Maldacena:2001ss, Aharony:1998qu, Witten:1998wy, Kravec:2013pua, Kravec:2014aza, Belov:2004ht}. A bulk theory that is equivalent to the \((p+1)\)-form Maxwell theory consists of the following Lagrangian:
\begin{equation}\label{eq:bulkmaxsf}
    S_{\rm bulk} = \int_{M_{d+1}} - \frac{1}{2 g^2} f_{p+2} \wedge \ast_{d+1} f_{p+2} + f_{p+2} \wedge d b_{d-p-2} \, .
\end{equation}
This Lagrangian leads to \((p+1)\)-form Maxwell theory when integrating out \(b_{d-p-2}\), since its equation of motion leads to \(d f_{p+2} = 0\). We can also dualize it directly to Maxwell theory for \(b_{d-p-2}\). In this case, we have the presence of a topological term \(f_{p+2} \wedge d b_{d-p-2}\). An important aspect of having the topological term is that it will dominate when we study the bulk theory close to the boundary, which coincides with the derivative counting \cite{Witten:1998xy, Maldacena:2001ss}. One can study the boundary conditions of this action, in particular by imposing \(f_{p+2}\) to be freely varying at the boundary, together with the equation of motion \(d f_{p+2} = 0\). This results in a free theory living on the boundary, which is called a singleton theory \cite{Maldacena:2001ss}. We can consider the action as it is, which will lead to a singleton theory that is not manifestly Lorentz invariant on \(M_d\), but it is self-dual under electromagnetic duality and contains both \(a_{p+1}\) and its dual field in \(d\) dimensions. We can also add, as in \cite{Maldacena:2001ss}, the following boundary terms:
\begin{equation}  \label{eq:bndrymax}
    S_d = \int_{M_d} - \frac{1}{2 g^2_b} f_{p+2} \wedge \ast_d f_{p+2} + f_{p+2} \wedge b_{d-p-2} \, ,
\end{equation}
where \(g_b\) is the coupling constant on the boundary. When we integrate out the \(b_{d-p-2}\) field in the bulk, the Neumann boundary condition becomes
\begin{equation} \label{eq:modNeu}
    \frac{1}{g^2} \ast_{d+1} da_{p+1} \Big|_{\partial M_{d+1}} = \frac{1}{g_b^2} d \ast_d d a_{p+1} \, .
\end{equation}
The topological term in \eqref{eq:bulkmaxsf} cancels the second one in the boundary action \eqref{eq:bndrymax} due to the flatness of \(f_{p+2}\) when integrating out \(b_{d-p-2}\). When we evaluate the bulk action \eqref{eq:bulkmaxsf}, the first term, provided that the modified Neumann boundary conditions \eqref{eq:modNeu} are satisfied, exactly reproduces a Maxwell theory on the boundary. This is what we consider as a dynamical gauging of symmetry. We can consider another type of gauging, i.e., with higher derivative terms. It would be relevant to understand what this corresponds to and how the bulk action modifies. We plan to study this in the future. Moreover, in 4 dimensions, we can add a topological theta term. This will correspond to adding the Chern-Simons term to the 3-dimensional boundary condition, as seen in \cite{DiPietro:2019hqe}.

\subsection{Sandwich Construction}
\label{sec:Sandwich}
In the sandwich picture for SymTFTs, the partition function of a TFT that lives on the manifold \(M_{d+1} = M_d \times I\), with a topological boundary condition at \(x = L\) and a QFT with a discrete symmetry that lives on the boundary at \(x = 0\), is shown to be equal to the partition function of a topological deformation of the boundary QFT. This is done by shrinking the interval \(I\) to zero, which is possible because the bulk theory is topological and does not depend on \(L\), \cite{Gaiotto:2014kfa, Pulmann:2019vrw, Freed:2022qnc}. The sandwich construction of the SymTFT also aims at capturing all topological manipulations related to a given symmetry $G$ of a quantum field theory. The SymTh does not only capture the topological manipulations, but also encodes symmetry operations that are not topological such as dynamical gauging. 

For the SymTh we are describing, the picture becomes more complicated. The Maxwell theory is not topological, and its partition function will depend on the length \(L\) of the interval \(I\). In this section, we will factorize the partition function, separating its dependence on the quantum fluctuations on the boundary. We will also provide a heuristic argument about the limit \(L \to 0\) and its result.

Let us start by considering the case with two Dirichlet boundary conditions for \(x = 0\) and \(x = L\). We can write these boundary conditions as states:
\begin{equation}
    \bra{D(\tilde{a}_{p+1,0})}_{x=0};\;\;
    \ket{D(\tilde{a}_{p+1,L})}_{x=L} \, .
\end{equation}
We see that the partition function of this theory is equal to the Euclidean propagator for Maxwell theory on the space manifold \(M_d\): \(G_d(\tilde{a}_{p+1,0}, \tilde{a}_{p+1,L}, L)\). We expect this propagator to become a delta function \(\delta(\tilde{a}_{p+1,0} - \tilde{a}_{p+1,L})\) as \(L \to 0\). We will now describe a construction to explain this limit.

If we fix the gauge, we will find one classical solution for every choice of \(\tilde{a}_{p+1,0}\) and \(\tilde{a}_{p+1,L}\). We will call such classical solutions \(a_{p+1,cl}\). We can then decompose any field configuration as the classical solution plus a fluctuation \(a_{p+1,\delta}\) that vanishes on the boundary. The action splits accordingly:
\begin{equation}\label{eq:sandwichfactor}
\begin{split}
    a_{p+1} &= a_{p+1,cl} + a_{p+1,\delta} \\
    S &= S_{cl}(a_{p+1,cl}) + S_\delta(a_{p+1,\delta}) \\
    Z &= e^{-S_{cl}} \int_{a_{p+1}|_{\partial M_{d+1}} = 0} D a_{p+1,\delta} \, e^{-S_\delta} \, .
\end{split}
\end{equation}
The partition function factorizes into the path integral over all the fluctuations in the bulk, which we will denote as \(Z_{\rm bulk}\), times the classical contribution \(e^{-S_{cl}}\). It will be useful to note that only part of the form \(a_{p+1}\) is fixed by the boundary condition, specifically \(a_{p+1}|_{\partial M}\), while \( *_{d+1}a_{p+1}|_{\partial M} \) is free on the boundary. We can use this last contribution to write a functional that is normalized to 1 on the space of classical configurations:
\begin{equation}\label{eq::ap+1cldelta}
    \frac{e^{-S_{cl}(a_{p+1,cl})}}{\int D a_{p+1,cl} e^{-S_{cl}(a_{p+1,cl})}} \, .
\end{equation}
notice how there is a single classical solution for every choice of boundary values $\tilde{a}_{p+1,0}$ and $\tilde{a}_{p+1,0}$. The integral at the denominator is over different boundary values, corresponding to different classical solutions

Now, as \(L \to 0\), the action for every classical solution with \(\tilde{a}_{p+1,0} \neq \tilde{a}_{p+1,L}\) tends to infinity. Therefore, it can be argued that \eqref{eq::ap+1cldelta} becomes a normalized functional with support only on \(\tilde{a}_{p+1,0} = \tilde{a}_{p+1,L}\), i.e., a delta function on the space of field configurations on \(M_d\):
\begin{equation}
    Z_{\rm bulk} \cdot e^{-S_{cl}} = \braket{D(\tilde{a}_{p+1,0})}{D(\tilde{a}_{p+1,L})} \sim_{L \to 0} \delta(\tilde{a}_{p+1,0} - \tilde{a}_{p+1,L}) \, .
\end{equation}
The same argument holds if we add terms to the action \eqref{eq:Max}. For example, we could add Chern-Simons terms to describe anomalies. 

We can take \(Z_{\rm bulk}\) as a working definition for the delta functional. A better interpretation of this procedure is that we have decoupled the bulk physics from the boundary by factorizing out \(Z_{\rm bulk}\) and dividing by the normalization factor in \eqref{eq::ap+1cldelta}.

Let us now consider the same theory with a Dirichlet boundary condition \(\ket{a_{p+1,L}}\) for \(x = L\) and a QFT with a \(U(1)\) symmetry and a background field \(a_{p+1,0} = a_{p+1}|_{x = 0}\) at \(x = 0\). We have a Maxwell action on \(M_{d+1}\), and the action of the QFT depends on several fields \(\phi_i\) on \(M_d\):
\begin{equation}
    S_{\rm tot} = S_{d+1}(a_{p+1}) + S_d(\phi_i, a_{p+1,0}) \, .
\end{equation}
The term \(S_d\) includes the coupling \eqref{eq:J-a_coupling}, and the gauge field \(a_{p+1}\) can also induce non-trivial holonomies on the boundary theory. The total partition function is:
\begin{equation}
    \begin{split}
        Z_{\rm tot} &= \int D a_{p+1} e^{-S_{d+1}(a_{p+1})} \int_{*_{d} J^{\rm QFT} = *_{d+1} da_{p+1}|_{\partial M}} \prod_i D \phi_i e^{-S_d(\phi_i, \tilde{a}_{p+1,0})} \\
        &= \int D a_{p+1} e^{-S_{d+1}(a_{p+1})} \cdot \tilde{Z}_d(\tilde{a}_{p+1,0}, \tilde{J}^{\rm QFT}) \, .
    \end{split}
\end{equation}
where the value of \(a_{p+1}|_{x=L}\) is fixed to \(\tilde{a}_{p+1,L}\), and \(\tilde{Z}_d(\tilde{a}_{p+1,0}, \tilde{J}^{\rm QFT})\) is the partition function for the boundary QFT in the presence of the background field \(\tilde{a}_{p+1,0}\), under the constraint \eqref{eq:constMNeumann}. If we integrate over \( *_{d+1} da_{p+1}|_{\partial M} \), we can obtain the unconstrained partition function:
\begin{equation}
\begin{split}
    &\tilde{Z}_d(\tilde{a}_{p+1,0}, \tilde{J}^{\rm QFT}) = \int_{*_{d} J^{\rm QFT} = *_{d+1} da_{p+1}|_{\partial M}} \prod_i D \phi_i e^{-S_d(\phi_i, \tilde{a}_{p+1,0})} \\
    &\int D \tilde{J}^{\rm QFT} \tilde{Z}_d(\tilde{a}_{p+1,0}, \tilde{J}^{\rm QFT}) = Z_d(\tilde{a}_{p+1,0}) \, .
\end{split}
\end{equation}
Now, as \(L \to 0\), the value of \(\tilde{a}_{p+1,0}\) is fixed to \(\tilde{a}_{p+1,L}\), while the integral over the unconstrained elements of the field \( *_{d+1} da_{p+1} \) becomes an integral over \(\tilde{J}^{\rm QFT}\). We can write the boundary state associated with the boundary theory as:
\begin{equation}
    \bra{QFT}_{x=0} = \int D \tilde{a}_{p+1,0} \bra{\tilde{a}_{p+1,0}} \cdot Z_d(\tilde{a}_{p+1,0}) \, .
\end{equation}
To perform the computation, we separate the integral over the boundary field \(\tilde{a}_{p+1,0}\) from the path integral. We can then write a classical solution in the bulk \(a_{p+1,cl}\) for every such configuration. The partition function becomes:
\begin{equation}
    \begin{split}
        \braket{QFT}{D(\tilde{a}_{p+1,L})} &= \int D \tilde{a}_{p+1,0} \braket{D(\tilde{a}_{p+1,0})}{D(\tilde{a}_{p+1,L})} \cdot Z_d(\tilde{a}_{p+1,0}) \, .
    \end{split}
\end{equation}
We used the same procedure as in \eqref{eq:sandwichfactor}, but this time we performed a path integral over all possible initial conditions \(\tilde{a}_{p+1,0}\). As \(L \to 0\), the classical action becomes a delta function, and we are left with \(Z_d(\tilde{a}_{p+1,L})\).

In the absence of Chern-Simons coupling potentially leading to anomalies for the field \( a_{p+1} \), we can instead impose Neumann boundary conditions at \( x = L \). To capture the dynamical gauging introduced by the boundary Maxwell action:
\[
S'_d = -\frac{1}{2g'^2} \int_{x=L} d\tilde{a}_{p+1} \wedge *_{d} d\tilde{a}_{p+1},
\]
we directly implement the modified Neumann boundary conditions introduced in equation \eqref{eq:modNeu}. We have seen that this can be related to the singleton sector when attempting to add the dualization topological term in the Lagrangian.

Now, this action exhibits a \( (d-p-3) \)-form symmetry described by the current \( *d\tilde{a}_{p+1} \). We can also introduce a background field \( \tilde{B}_{d-p-2} \) associated with this symmetry. The total action becomes:
\[
\begin{split}
    S_{\text{tot}} &= S_d(\phi_i, a_{p+1,0}) + \int_{M} \left( -\frac{1}{2g^2} f_{p+2} \wedge *_{d+1} f_{p+2} + \tilde{B}_{d-p-2} \wedge f_{p+2} \right) \\
    &\quad + \int_{x=L} \left( -\frac{1}{2g'^2} f'_{p+2} \wedge *_{d} f'_{p+2} + (d\tilde{B}_{d-p-2} + b_{d-p-2}) \wedge f'_{p+2} \right) \\
    &= S_d(\phi_i, \tilde{a}_{p+1,0}) + S_{d+1}(f_{p+2}, b_{d-p-2}) + S_G(f'_{p+2}, \tilde{B}_{d-p-2}, b_{d-p-2}),
\end{split}
\]
where \( S_G \) denotes the boundary action. We impose the modified boundary condition on \( b_{d-p-2} \) to set \( f_{p+2} = f'_{p+2} \), then integrate out the field, yielding \( f_{p+2} = d a_{p+1} \). 

Taking the limit \( L \to 0 \), as before, we have \( a_{p+1}|_{x=0} = a_{p+1}|_{x=L} = \tilde{a}_{p+1} \). In this case, both \( \tilde{a}_{p+1} \) and \( *_{d+1}da_{p+1}|_{x=0} = *_{d} J^{\text{QFT}} \)  are integrated over all possible configurations. The left boundary state becomes:
\[
\ket{N(\tilde{b}_{d-p-2})}_{x=L} = \int D\tilde{a}_{p+1,L} \, e^{-S_G(\tilde{a}_{p+1,L}, \tilde{b}_{d-p-2})} \, \ket{\tilde{a}_{p+1,L}}.
\]

The partition function is then given by:
\[
\begin{split}
    \braket{QFT}{N(\tilde{b}_{d-p-2})} &= \int D \tilde{a}_{p+1,0} D \tilde{a}_{p+1,L} \, \braket{D(\tilde{a}_{p+1,0})}{D(\tilde{a}_{p+1,L})} \cdot Z_d(\tilde{a}_{p+1,0}) \cdot e^{-S_G(\tilde{a}_{p+1,L}, \tilde{b}_{d-p-2})}.
\end{split}
\]
Taking the limit \( L \to 0 \) again results in a delta function, and the partition function simplifies to:
\[
\braket{QFT}{N} = \int D \tilde{a}_{p+1,L} \, Z_d(\tilde{a}_{p+1,L}) \cdot e^{-S_G(\tilde{a}_{p+1,L}, \tilde{b}_{d-p-2})}.
\]
This represents the partition function of the QFT coupled to a dynamical field \( a_{p+1,L} \) with Maxwell action and background field \( \tilde{b}_{d-p-2} \).

Notice that the action of the topological operators, as described in equation \eqref{eq:action_on_Wsurf}, does not depend on \( L \). Thus, although the limit \( L \to 0 \) provides valuable information about the theory we are describing, we do not need to explicitly take this limit to study the topological operators of a given QFT.

The procedure developed in this subsection can be generalized to compute bulk correlators and other quantities by factorizing the contribution of the fluctuations in the bulk.

\subsection{Truncation to Symmetry TFT}
Finally, let us comment on the relation between our approach and the one taken in \cite{Brennan:2024fgj, Antinucci:2024zjp, Bonetti:2024cjk}. For instance, it is possible to rewrite the Maxwell action using the Lagrange multiplier \( h_{d-p-2} \):
\begin{equation}
    S_{d+1} = -\frac{g^2}{2} \int_{M_{d+1}} v \; h_{d-p-2} \wedge \ast h_{d-p-2} + \int_{M_{d+1}} h_{d-p-2} \wedge d a_{p+1},
\end{equation}
where \( v = (-1)^{(p+3)(d-p-2)} s \) and \( s \) is the parity of the signature of the metric. It is straightforward to show that by integrating out \( h_{d-p-2} \), we recover the \( (p+1) \)-form Maxwell action \eqref{eq:Max}. If we instead truncate the theory to the topological term, \( \int_{M_{d+1}} h_{d-p-2} \wedge d a_{p+1} \), the field \( h_{d-p-2} \) acquires an extra gauge invariance, \( h_{d-p-2} \to d \lambda_{d-p-3} \), becoming an \( \mathbb{R} \)-valued gauge field \footnote{It is worth stressing that the strict $g=0$ limit of Maxwell theory does not coincide with the usual infrared limit $g \rightarrow
 0$. In the deformed BF parametrization, the latter limit has to be taken scaling $F \sim g^2$ not to suppress photon propagation. This results in a gapless theory of free photons, while the strict $g=0$ limit in a gapped theory.}. This is exactly the BF topological action considered in \cite{Brennan:2024fgj, Antinucci:2024zjp}. The non-abelian counterpart of this topological action, as the strict \( g = 0 \) limit of Yang-Mills, has also been discussed in \cite{Bonetti:2024cjk}. In 2D, the relation between Maxwell and Yang-Mills, and their topological BF counterpart, is recovered in the 0-volume limit, which is equivalent to taking the strict \( g = 0 \) limit. This can be seen by computing the partition function of the 2D theory.

\section{Lower-dimensional examples\label{eq:ldex}}

In this section, we consider the special case of \( d = 1 \) of our general construction, where the bulk is two-dimensional. In this setting, only 0-form symmetries can be discussed. Thus, compared to \autoref{sec:setup}, we drop all form and dimension indices and indicate the gauge connection and its curvature by \( A \) and \( F \), respectively. Two-dimensional gauge theories have been extensively studied in the past \cite{Blau:1991mp, Witten:1992xu, Witten:1991we} and are known for their remarkable simplicity and almost-topological character. For the sake of our work, they represent an ideal testing ground that does not stray too far from the original SymTFT ideas.

For 2D Maxwell theory, the topological operators are
\begin{equation}
    U_{e^{i \alpha}}^e(x) = \exp\left\{\frac{\alpha}{g^2} (\ast F)(x) \right\}, \quad U^m_{e^{i \alpha}}(\Sigma_{b,g}) = \exp\left\{ \frac{i \alpha}{2\pi} \int_{\Sigma_{b,g}} F \right\} \, .
\end{equation}
The magnetic operator is trivial in 2D: no operators are charged under it, and its insertion in partition functions only amounts to a shift in the \( \theta \)-angle for 2D Maxwell theory. We will discuss the possible known generalization to the case of 2D Yang-Mills in the dedicated section below. Finally, Llne operators charged under \( U^e \) are ordinary Wilson lines $W_q$.


\subsection{Boundary conditions in \( d = 2 \)}

Let us comment on some peculiarities regarding boundary conditions for \( 2d \) Maxwell theory (a generalization of the discussion to \( 2d \) Yang-Mills is straightforward). In \( d = 2 \), \( F \) has only one gauge-invariant component, and a convenient way to impose a boundary condition at some one-dimensional boundary \( \gamma_i \) is to fix the holonomy of the connection to some group element \( g_i \) by inserting in path integrals
\begin{equation}\label{eq:Hol_delta}
    \delta\left( e^{i \int_{\gamma_i} A} , g_i \right) = \sum_{n \in \mathbb{Z}} e^{i n \alpha_i} e^{- in \int_{\gamma_i} A}, \quad g_i = e^{i \alpha_i}  \, .
\end{equation}
This boundary condition does not have a direct analogue in higher dimensions for one-form gauge fields. However, it is related to the Dirichlet boundary condition in a precise way, as we now comment. Since in $d=2$ all $U(1)$ bundles restrict to trivial bundles on a $d=1$ boundary $\gamma_i$ (with necessarily flat connections as $F|_{\gamma_i} =0$), the only gauge-invariant information is contained in boundary holonomies. This information is also encoded in the non-exact part of boundary connections:
\begin{equation}
A|_{\gamma_i} =  \alpha_i \frac{\dd \theta}{2\pi} +  \Omega^0(S^1), \quad \alpha \sim \alpha + \mathbb{Z} \, .
\end{equation}
An alternative way to impose such boundary condition is via variational principle with the action
\begin{equation}\label{eq:alternative-Hol}
S_{\partial} =  i \int_{\gamma_i} \phi \left( A - \alpha_i \frac{\dd \theta}{2\pi} - \dd \lambda \right)  
\end{equation}
where $\phi$ is an auxiliary field integrating to on $\gamma_i$ to $2\pi \mathbb{Z}$ and $\lambda \in \Omega^0(S^1)$ is a St\"uckelberg field. This action is gauge invariant under small gauge transformations as well as for large ones parametrized by $\Lambda = k \dd \theta$, for which
\begin{equation}
\delta_\Lambda S_\partial = i k \int \phi \dd \theta = 2 \pi i \mathbb{Z} \, .
\end{equation}
Moreover, it leads to the following boundary equations of motion:
\begin{equation}
\frac{1}{g^2} \ast F = i \phi, \quad A = \alpha_i \frac{\dd \theta}{2\pi} + \dd \lambda, \quad \dd \phi = 0 \, ,
\end{equation}
which make manifest its equivalency with the fixed-holonomy boundary conditions. It also shows that Wilson lines $W_q$ are allowed to terminate on operators of the form $e^{ iq \lambda}$ and that electric topological operators are non-trivial: this is what is expected from a Dirichlet boundary. 

If we integrate out $\phi$ by solving its equation on motion, this constraints $\phi \in \mathbb{Z}$. The resulting partition function on a single-boundary surface, say, a disk $D$ with $\partial D= \gamma$, reads
\begin{equation}
\mathcal{Z}(D,e^{i \alpha}) = \sum_{n \in \mathbb{Z}} e^{i n \alpha} \int \mathcal{D} A \, e^{\frac{1}{2g^2} \int F \wedge \ast F - i n \int_\gamma A}.
\end{equation}
which is precisely the expression one finds using the alternative definition of the holonomy boundary condition \eqref{eq:Hol_delta}. This latter formula has an alternative interpretation which connects to the other relevant boundary condition for Maxwell theory, the Neumann boundary condition. In fact, the path integrals over sectors labeled by \( n \in \mathbb{Z} \) all lead to a well-defined variational problem with boundary condition \( \ast F|_{\gamma} =  i n g^2 \). The \( n = 0 \) term precisely corresponds to a Neumann condition, while the \( n \neq 0 \) terms correspond to a modified Neumann condition, \( MN_n(A) \), in which the bulk field \( A \) is coupled to a constant background current \( J_1 = n \dd \theta \). In terms of boundary states, we can then write
\begin{equation}
| \text{Hol}_\gamma(A) = e^{i \alpha} \rangle = | N(A) \rangle + \sum_{n \neq 0} e^{i n \alpha} | MN_{n}(A) \rangle \,.
\label{eq:holbc}
\end{equation}
As we will comment later, the limit $\alpha \rightarrow 0$ corresponds to closing a boundary. In the specific case of the disk, the integer \( n \in \mathbb{Z} \) can then be identified with the monopole number for the \( U(1) \)-bundles one obtain on \( S^2 \). 

The purely Neumann boundary condition can be implemented modifying \eqref{eq:alternative-Hol} by (flat) gauging the $U(1)$-symmetry implemented on the boundary holonomy shifting $\alpha_i$. It is easy to see that integrating out the auxiliary fields this will lead to the variational problem in the absence of any extra boundary action $S_\partial$, with boundary equation of motion $\ast F |_{\gamma_i}= 0$.  

We will not be interested in what follows in Neumann boundary conditions, as they will corresponds to symmetry boundaries for $\mathbb{Z}^{(-1)}$ symmetries as their topological operators (the Wilson lines on the boundary) are space-filling. Instead, we will exclusively focus on holoomy-fixing (Dirichlet) boundary conditions.

\subsection{Aspects of $2d$ Maxwell theory}

Gauge theories in $d=2$ are expecially simple due to their almost topological character. For example, the partition function of \( \text{Maxwell}_2 \) on a Riemann surface \( \Sigma_{b,g} \) with \( b \) boundary curves \( \gamma_{i=1,...,b} \) with holonomy-fixing boundary conditions and genus \( g \) 
\begin{equation}
\mathcal{Z}(\Sigma_{b,g} \, | \, g_1, \dots, g_b) = \int \mathcal{D} F \, \exp\left\{ - \frac{1}{2g^2} \int_{\Sigma_{b,g}} F \wedge \ast F  + \frac{i \theta}{2\pi} \int F \right\} \left. \prod_{i=1}^b \delta\left( e^{i \int_{\gamma_i} A}, g_i \right) \right|_{\rm gauge\, fixed} \, .
\end{equation}
Can be computed in closed form \cite{Migdal:1975zg,Rusakov90,Witten:1991we,blau1992quantum}. Adapting the more general results for \( \text{YM}_2 \) to the simpler representation theory of \( U(1) \), one finds
\begin{equation}
\mathcal{Z}(\Sigma_{b,g} \, | \, g_1, \dots, g_b) = \sum_{n \in \mathbb{Z}} \left[ \prod_{i=1}^b e^{i n \alpha_i} \right] e^{- \frac{g^2}{2} \mathcal{A}(\Sigma_{b,g}) \left( n - \frac{\theta}{2\pi} \right)^2}, \quad g_i = e^{i \alpha_i} \, ,
\end{equation}
where \( \mathcal{A}(\Sigma_{b,g}) \) is the area of the surface \( \Sigma_{b,g} \). This result makes manifest the fact that \( \text{Maxwell}_2 \) is almost topological: it is invariant under area-preserving diffeomorphisms. The topological limit is obtained by tuning to zero the only dimensionless ratio of the theory, \( g^2 \mathcal{A} \rightarrow 0 \).

Similarly, partition functions with insertions of Wilson lines and loops can be obtained using specific cutting and gluing rules. For example, the disk partition function with two \( q \)-Wilson lines joining two points on the boundary is:
\begin{equation}\label{eq:WilsonLoops}
\mathcal{Z}(D , W_q , W_q | e^{i \alpha_1}) = \sum_{n \in \mathbb{Z}} e^{i \left( n + q \right) \alpha_1} e^{- \frac{g^2}{2}\left[ \mathcal{A}(D^{(\rm in)}) \left( n - \frac{\theta}{2\pi} \right)^2 + \mathcal{A}(D^{(\rm out)}) \left(n+q- \frac{\theta}{2\pi}\right)^2 \right]} \,
\end{equation}
where \( D^{(\rm in)} \) is the disk topology enclosed by the two lines and \( D^{(\rm out)} \) is the disk area outside the lines. The same result for the cylinder topology \( C \) can be obtained by \( \alpha_1 \rightarrow \alpha_1 - \alpha_2 \), which is equivalent to adding a boundary with opposite orientation and holonomy fixed at \( g_2 = e^{i \alpha_2} \). On the cylinder, one can have the same configuration where one line winds around the compact direction once: the result is identical, but the two areas have to be identified with the areas of the surfaces enclosed by the two lines. This is consistent with the fact that there is no limit to which the internal area can shrink. In what follows, we will neglect the \( \theta \)-term for convenience, as it is straightforward to reinstate it in our formulas. 

We will also consider Wilson lines unattached to holonomy-fixing boundaries. This allows to use closed form expressions such as \eqref{eq:WilsonLoops} without passing through the variational principle of \eqref{eq:Hol_delta} when Wilson lines with endpoints are included. Since Wilson loops insertions are invariant under area-preserving deformations, this does not affect the result of the slab compactification limit. 

In the rest of the section, we explore the possibility of using \( \text{Maxwell}_2 \) as a SymTh for a quantum mechanics (QM) system living on the boundary \( \gamma_1 \) of a cylinder topology with area \( \mathcal{A} \). On the second boundary \( \gamma_2 \), we provide a boundary condition in terms of the gauge field holonomy. All we require for the QM theory is to have a 0-form \( U(1) \)-global symmetry with current \( J \) and local operators charged under it. The corresponding Ward identity is:
\begin{equation}
\label{eq:WI}
\langle (\dd \ast_1 J)(t) \prod_{k=1}^n \mathcal{O}_k(t_k) \rangle = - i \sum_{\alpha = 1}^n q_\alpha \delta(t - t_\alpha) \langle \prod_{k=1}^n \mathcal{O}_k(t_k) \rangle \dd t, \quad q_\alpha \in \mathbb{Z}.
\end{equation}
This \( U(1) \)-global current is coupled to the bulk gauge field, which effectively defines an interacting boundary condition. The partition functions of the joint system with boundary operator insertions are:
\begin{align}
\label{eq:fullpartition}
\mathcal{Z}_{\text{Maxwell}_2 + \text{QM}} &= \int \mathcal{D} F \, \exp\left\{ - \frac{1}{2g^2} \int F \wedge \ast F  + \frac{i \theta}{2\pi} \int F \right\} \left. \delta\left( e^{i \int_{\gamma_2} A}, g_2 \right) \right|_{\rm gauge\, fixed} \nonumber\\
&\times \mathcal{N}(t_1, \dots, t_m) \langle\, e^{i \int A \wedge \ast_1 J} \prod_{i=1}^m \mathcal{O}_{q_i}(t_i) \,\rangle_{\rm QM} \, ,
\end{align}
where \( \langle \dots \rangle_{\rm QM} \) indicates averages in the QM theory, and \( \mathcal{N}(t_1, \dots, t_m) \) represents a network of appropriately chosen Wilson lines extending in the bulk, with endpoints at the boundary operator insertions. This term is necessary to achieve gauge invariance when charged operators are inserted at the boundary. Since the QM at the boundary is not topological in general, one cannot move the insertions \( t_1, t_2, \dots \) freely.

We will also consider the insertion of topological operators \( U_g^e \) in this setup. A pictorial representation of a generic configuration of insertions is:
\begin{equation}
\mathcal{Z}_{\text{Maxwell}_2 + \text{QM}} = \vcenter{\hbox{\includegraphics[scale=0.8]{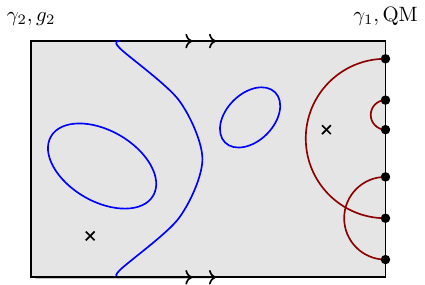}}}
\end{equation}
where black dots indicate insertions of QM operators \( \mathcal{O}_q \), crosses represent insertions of the topological operator \( U_g^e \), and red and blue lines represent Wilson lines and Wilson loops, respectively. Wilson lines joining boundary operator insertions are allowed to wind around the cylinder or meet at the same point on the opposite boundary. We do not indicate the orientation of the line operators, as it can be deduced on a case-by-case basis following gauge invariance.

To compute these partition functions, we can make use of a feature special to the QM system: the Ward identity \eqref{eq:WI} implies that the current \( \ast_1 J \), inserted in a correlator with charges \( q_i \) at positions \( t_1 < t_2 < \dots < t_k \), can be written as a piecewise function:
\begin{equation}
\ast_1 J = q_1 I_{[t_1,t_2]} + (q_2 + q_1) I_{[t_2,t_3]} + (q_3 + q_2 + q_1) I_{[t_3,t_4]} + \dots = \sum_{\alpha=1}^k \left( \sum_{i=1}^\alpha q_i \right) I_{[t_\alpha,t_{\alpha+1}]} \, ,
\end{equation}
where \( I_{[t_1,t_2]} = \theta(t - t_1) \theta(t_2 - t) \). In particular, in our setting we have:
\begin{equation}
\langle e^{i \int_{\gamma_1} A \wedge \ast_1 J} \prod_{i=1}^k \mathcal{O}_{q_k}(t_k) \rangle = e^{i q_1 \int_{\gamma_{1,[t_1,t_2]}} A} e^{i (q_2+q_1) \int_{\gamma_{1,[t_2,t_3]}} A} \dots \times \langle  \prod_{i=1}^k \mathcal{O}_{q_k}(t_k) \rangle \, .
\end{equation}
This allows us to factorize a partition function with generic configurations as:
\begin{equation}
\mathcal{Z}_{\text{Maxwell}_2 + \text{QM}} = \bar{\mathcal{Z}}_{\text{Maxwell}_2} \times \langle \prod_{i=1}^k \mathcal{O}_{q_k}(t_k) \rangle \, ,
\end{equation}
where \( \bar{\mathcal{Z}}_{\text{Maxwell}_2} \) is the partition function we started with, but with operator insertions replaced by a sequence of Wilson lines living on the boundary \( \gamma_1 \), as expressed above. This new partition function carries information about the boundary insertions through the endpoints of the Wilson lines and their charge.

In the interval compactification limit, the \( \mathcal{O}_{q_k} \) correspond to non-genuine operators of the QM charged under the \( U(1) \). As a worked-out example, consider the partition function \( \bar{\mathcal{Z}}_{\text{Maxwell}_2} \) obtained by inserting the operators \( \mathcal{O}_q(t_1), \, \mathcal{O}_q(t_2) \) on the boundary, with two insertions of topological operators and a Wilson line joining them, winding around the cylinder. In our graphical representation, this is:
\begin{equation}
\label{eq:myexample}
\vcenter{\hbox{\includegraphics[scale=0.8]{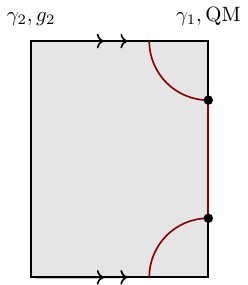}}} = e^{i q \alpha_2} e^{- \frac{g^2}{2} \mathcal{A} q^2} \langle \mathcal{O}_q(t_1) \mathcal{O}_{-q}(t_2) \rangle
\end{equation}

Insertion of topological operators in the cylinder partition function with no QM coupling and boundary conditions specified by \( g_{1,2} = e^{i \alpha_{1,2}} \) results in:
\begin{equation}
	\langle \prod_{i = 1}^k U_{\alpha_i}(x_i) \rangle_C  = \sum_{n \in \mathbb{Z}} e^{in(\alpha_1 - \alpha_2)} e^{- \frac{g^2}{2} \mathcal{A}(C) n^2} \prod_{i=1}^k e^{i n \alpha_i} \, ,
\end{equation}
which shows that the topological operators implement a \( U(1) \)-action on the boundary states. Their action on a Wilson loop \( W_q \) is, as expected, given by:
\begin{equation}
\vcenter{\hbox{\includegraphics[scale=0.8]{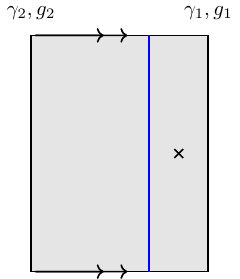}}} = e^{i q \alpha} \vcenter{\hbox{\includegraphics[scale=0.8]{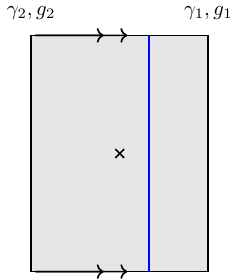}}} 
\end{equation}
When this relation is applied to our toy computation in equation \eqref{eq:myexample}, with the insertion of topological operators, one obtains:
\begin{equation}
\label{eq:symmop}
\vcenter{\hbox{\includegraphics[scale=0.8]{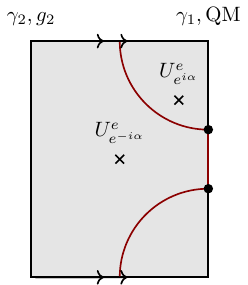}}} = e^{i q \alpha} e^{i q \alpha_2} e^{-\frac{g^2}{2} A q^2} \langle \mathcal{O}_q(t_1) \mathcal{O}_{-q}(t_2)\rangle
\end{equation}
This result reproduces the transformation law of the operator \( \mathcal{O}_q(t) \) under the action of its global symmetry topological operator:
\begin{equation}
U_{e^{i \alpha}}(t^-) \mathcal{O}(t) U_{e^{i \alpha}}^{-1}(t^+) = e^{i q \alpha} \mathcal{O}(t), \quad U_{e^{i\alpha}}(t) = \exp\left\{ i \alpha (\ast_1 J)(t) \right\} \, .
\end{equation}
In simpler terms, this follows from the fact that the interaction of the QM with the bulk fields on \( \gamma_1 \) imposes a general boundary condition which identifies the bulk topological current and the boundary global current: \( \ast F |_{\gamma_1} = \ast_1 J \).

To conclude, \( \text{Maxwell}_2 \) can be used to capture the symmetry properties of a QM theory living on a boundary exactly. Due to its simplicity, it has not been necessary to take the limit \( \mathcal{A} \to 0 \) or weak coupling \( g \to 0 \), which are identified for dimensional reasons. In the topological limit, all area factors from the results above disappear, and objects of the form \eqref{eq:fullpartition} exactly compute symmetry properties of QM correlators.

\subsection{$2d$ Yang-Mills theory}

Yang-Mills theory in two dimensions is also invariant under area-preserving diffeomorphisms, and its partition function can be computed exactly on any surface \( \Sigma_{g,b} \):
\begin{equation}
\mathcal{Z}(\Sigma_{g,b}, g_1, \dots, g_b) = \sum_\lambda \text{dim}(\lambda)^{2-2g-b} \left[\prod_{i=1}^b \chi_\lambda(g_i) \right] e^{- \frac{g^2}{2} A(\Sigma_{g,b}) c_2(\lambda)} \, ,
\end{equation}
where \( \lambda \) are irreducible representations of the gauge group \( G \), and \( \chi_\lambda \) is their character. On the cylinder, one can also compute the average of Wilson loops or pairs of Wilson lines joining the two boundaries. For example, the average of a Wilson loop wrapping around the cylinder cycle is:
\begin{align}
	\langle \mathcal{W}_\mu[\gamma] \rangle_C &=  \sum_{\lambda, \rho} N_{\lambda \mu}^\rho \chi_\lambda(g_1) \chi_\rho(g_2) e^{- \frac{g^2}{2} \left[ A(C^{(1)}) c_2(\lambda) + A(C^{(2)}) c_2(\rho) \right]}  \, ,
\end{align}
where \( N_{\lambda \mu}^\rho \) is the multiplicity of the representation \( \rho \) in the product \( \lambda \otimes \mu \), and \( C^{(1,2)} \) are the two sides of the cylinder. 

When we add a quantum mechanical (QM) system with a \( G \)-symmetry group on one boundary, we can again build gauge-invariant partition functions, as in equation \eqref{eq:fullpartition}, and verify whether these observables capture the symmetry properties of the boundary QM. 

The main challenge lies in defining an analogue of the \( U_g^e \) symmetry operators from two-dimensional Maxwell theory. These should be replaced by Gukov-Witten operators \cite{Gukov:2008sn, Gukov:2014gja}, for which a two-dimensional version and some of their properties have been studied in \cite{Nguyen:2021naa}. These operators, \( V_{[U]}(x) \) with \( U \in G \), are defined over conjugacy classes of \( G \) as disorder operators: remove a point \( x \in \Sigma_{b,g} \), and compute the path integral over gauge connections with boundary conditions on the holonomy \( \text{hol}_{\gamma_x}(A) \in [U] \), where \( \gamma_x \) is a small closed curve surrounding \( x \). The authors of \cite{Nguyen:2021naa} also show that these operators act non-invertibly on contractible Wilson loops, while only those operators valued in the center of \( G \) act in an invertible fashion. 

If we impose the holonomy boundary condition, we expect to restore the full symmetry \( G \) at the boundary in the interval compactification. In other words, the Gukov-Witten operators split into the individual elements of the conjugacy class under the Dirichlet boundary condition \cite{Antinucci:2022vyk, Antinucci:2022cdi}.

\section{Higher-dimensional examples\label{sec:hdex}} 
In this section, we focus on the symmetry theory for two different types of 2-groups. The first is the continuous 2-group in 4d, as analyzed in \cite{Cordova:2018cvg}.

\subsection{Continuous 2-group in 4d}
The bulk theory of a continuous 2-group in 4d has already been proposed in \cite{Damia:2022bcd}. Here, we review this theory and apply our prescription to this example. 

We are interested in studying the behavior of the bulk theory, which consists of two 1-form gauge fields interacting via a Chern-Simons term:
\begin{equation}
    S = \int_{M_{4+1}} \left( -\frac{1}{2} da_1 \wedge * da_1 - \frac{1}{2} dc_1 \wedge * dc_1 + \frac{k}{24\pi^2} a_1 \wedge dc_1 \wedge dc_1 \right).
\end{equation}
To compute the variation of the action for a manifold with boundary \( \partial M_{4+1} = M_4 \), we obtain:
\begin{equation}
    \begin{split}
        \delta S = & - \frac{1}{2} \int_{M_{4+1}} \delta a_1 \wedge d * da_1 - \frac{1}{2} \int_{M_{4+1}} \delta c_1 \wedge d * dc_1 \\
        & + \frac{1}{2} \int_{M_4} \delta a_1 \wedge *_{4+1} da_1 + \frac{1}{2} \int_{M_4} \delta c_1 \wedge *_{4+1} dc_1 \\
        & + \frac{k}{24\pi^2} \left( \int_{M_{4+1}} \delta c_1 \wedge da_1 \wedge dc_1 + \int_{M_{4+1}} \delta a_1 \wedge dc_1 \wedge dc_1 \right) \\
        & - \frac{k}{12\pi^2} \int_{M_4} \delta c_1 \wedge a_1 \wedge dc_1.
    \end{split}
\end{equation}
The equations of motion are then given by:
\begin{equation}
\begin{split}\label{eq:EOM_2group_U1}
    & d * da_1 = \frac{k}{12\pi^2} dc_1 \wedge dc_1, \\
    & d * dc_1 = \frac{k}{12\pi^2} da_1 \wedge dc_1.
\end{split}
\end{equation}
It is evident that if we impose Dirichlet boundary conditions for both \( a_1 \) and \( c_1 \), we constrain gauge transformations to vanish at the boundary. Alternatively, we might consider the following boundary conditions:
\begin{equation}
\begin{split}
    & *_{4+1} da_1|_{\partial M_{4+1}} = 0, \\
    & (*_{4+1} dc_1 - a_1 \wedge dc_1)|_{\partial M_{4+1}} = 0.
\end{split}
\end{equation}
However, the second equation is not invariant under gauge transformations that act on the boundary value of \( a_1 \). We can resolve this issue by setting Dirichlet boundary conditions for \( a_1 \), which eliminates such gauge transformations, or by setting Dirichlet boundary conditions for \( c_1 \), which removes the second equation.

Next, we examine the states described by the various boundary conditions. If we set two Dirichlet boundary conditions, we obtain a state that describes a theory with two \( U(1) \) symmetries with a mixed anomaly:
\begin{equation}
\begin{split}
    \braket{QFT}{D(\tilde{a}_{1}), D(\tilde{c}_{1})} &= Z_{DD}(\tilde{a}_{1}, \tilde{c}_{1}), \\
    Z_{DD}(\tilde{a}_{1} + d\lambda, \tilde{c}_{1}) &= Z(\tilde{c}_{1}, \tilde{c}_{1}) e^{\frac{k}{24\pi} \int_{M_4} \lambda dc_1 \wedge dc_1}.
\end{split}
\end{equation}
Alternatively, we can use Neumann boundary conditions for \( a_1 \). In this case, the theory has a non-invertible symmetry, arising from an ABJ anomaly (non-conservation law of the current), as described by the first equation of \eqref{eq:EOM_2group_U1}. This leads to the non-invertible bulk operators described in \cite{Damia:2022bcd}:
\begin{equation}
    \mathcal{D}_{1/N}^{(1,a)}(M_3) = \int [Da] \exp \left[ 2 \pi i \oint_{M_3} \left( \frac{1}{N} * J_2^{a} + \frac{N}{2} a \wedge da + a \wedge dc_1 \right) \right],
\end{equation}
where \( J_2^{a} = da_1 \). This operator can be projected parallel to the boundary, since the charged Wilson surface \( W_{a_1}(M_1) = e^{i \oint_{M_1} a_1} \) can end on the boundary, becoming a faithful 0-form non-invertible symmetry of the boundary QFT. The sandwich and interval compactification leads to:
\begin{equation}
\begin{split}
    \braket{QFT}{N, D(\tilde{c}_{1})} &= Z_{N,D}(\tilde{c}_{1}).
\end{split} 
\end{equation}
Notice that we didn't add a background gauge field \( \tilde{b}_2 \) due to the anomaly.

Finally, if we choose Dirichlet boundary conditions for \( a_1 \) and Neumann boundary conditions for \( c_1 \), we obtain a theory with a 2-group symmetry:
\begin{equation}
\begin{split}
    \braket{QFT}{D(\tilde{a}_{1}), N(\tilde{b}_{2})} &= Z_{DN}(\tilde{a}_{1}, \tilde{b}_{1}), \\
    Z_{DN}(\tilde{a}_{1} + d\lambda, \tilde{b}_{2} - \frac{k}{24\pi} \lambda d\tilde{a}_1) &= Z(\tilde{a}_{1}, \tilde{b}_{1}).
\end{split}
\end{equation}
To show this, we have added a counterterm that modifies the Chern-Simons term in the Lagrangian to \( \frac{k}{24\pi} c_1 \wedge da_1 \wedge dc_1 \), without altering the boundary conditions. It is interesting to note here that the background field $b$ is not a dynamical field appearing in the Bulk of our SymTh, but a field with  fixed value modifing the boundary conditions.

\subsection{Mixed case: 2-group with a discrete 1-form symmetry}
We now focus on a 2-group symmetry that mixes a discrete 1-form symmetry with a non-simply connected 0-form symmetry \cite{Benini:2018reh, Apruzzi:2021vcu, Apruzzi:2021mlh, Genolini:2022mpi, Bhardwaj:2022dyt, Mekareeya:2022spm, Carta:2022fxc, Lee:2021crt, Hsin:2020nts, DelZotto:2022joo, Cvetic:2022imb}. Specifically, we consider the following equation:
\begin{equation} \label{eq:2greq}
    \delta B_2 = Bock(A*w_2),
\end{equation}
where \( B_2 \) is the background for a discrete 1-form symmetry with \( \Gamma^{(1)} = \mathbb{Z}_N^{(1)} \), and \(A* w_2 \) is the pullback of the Brauer class of a \( G^{(0)} = PSU(N) \) global symmetry that obstructs the lift of \( PSU(N) \cong SU(N)/\mathbb{Z}_N \) to \( SU(N) \) bundles. The homomorphism \( Bock \) is associated with the sequence:
\begin{equation}
    0 \rightarrow \mathbb{Z}_N \rightarrow \mathbb{Z}_{N^2} \rightarrow \mathbb{Z}_N \rightarrow 0,
\end{equation}
and in particular, it is the map between cohomologies:
\begin{equation}
 Bock: H^2(M_{d+1}, \mathbb{Z}_N) \rightarrow H^3(M_{d+1}, \mathbb{Z}_N).
\end{equation}

We propose the bulk description as a mixture of BF-theory and Yang-Mills theory for the 0-form symmetry:
\begin{equation}
\begin{aligned}
    S_{\rm bulk} = & \int_{M_{d+1}} \left( -\frac{1}{2 g^2_{\rm YM}} \text{tr} \left( (f' - c_2 \mathbb{I}_N) \wedge * (f' - c_2 \mathbb{I}_N) \right) + N c_2 d \tilde{c}_{d-2} + N b_2 d \tilde{b}_{d-2} \right. \\
    & \quad \left. + d c_2 \, \tilde{b}_{d-2} + u \wedge (\text{tr}(f') - N c_2) \right),
\end{aligned}
\end{equation}
where \( u \) is a Lagrange multiplier that enforces \( \text{tr}(f') = N c_2 \) \cite{Seiberg:2018ntt, Gaiotto:2017yup}, with the trace taken in the fundamental representation. Additionally, we have:
\begin{equation}
    Bock(A*w_2) = \frac{d c_2}{N} \; \text{mod} \; N,
\end{equation}
where \( c_2 \) is the integral lift of \( w_2 \) \cite{Brennan:2023mmt}. The field strength \( f' = d a' \) corresponds to a \( U(N) \) gauge field, and the \( U(1) \in U(N) \) is reabsorbed by the gauge transformation:
\begin{equation}
    a' \rightarrow a' + \lambda, \qquad c_2 \rightarrow c_2 + d\lambda.
\end{equation}

By integrating out the Lagrange multiplier, we obtain:
\begin{equation}
\begin{aligned}
    S_{\rm bulk} = & \int_{M_{d+1}} \left( \frac{1}{2 g^2_{\rm YM}} \left( - \text{tr}(f' \wedge * f') + N c_2 \wedge * c_2 \right) + N c_2 d \tilde{c}_{d-2} + N b_2 d \tilde{b}_{d-2} \right. \\
    & \quad \left. + d c_2 \, \tilde{b}_{d-2} \right).
\end{aligned}
\end{equation}

In this formulation, we have decoupled the continuous \( SU(N) \) part from the discrete gauge fields, remembering that the \( U(1) \in U(N) \) is canceled by the transformation of the \( c_2 \) field. As usual, we place the quantum field theory (QFT) at \( x = 0 \). We will not discuss the topological operators of the bulk Yang-Mills theory, but will focus on the bulk gauge field, which has Dirichlet boundary conditions at \( x = L \), so that \( \mathfrak{su}(N) \) becomes the flavor algebra after compactifying the interval.

The global structure is dictated by the gapped boundary condition for the discrete topological part of the symmetry theory, in particular by the field \( c_2 \), which satisfies the standard finite symmetry TFT rules. The 2-group equation follows from the equations of motion of the finite part:
\begin{equation}
    d b_2 = \frac{d c_2}{N} \; \text{mod} \; N,
\end{equation}
which, with discrete fields, gives equation \eqref{eq:2greq}. The continuous part takes care of the relation between \( w_2 \) and \( c_2 \), and determines when \( \beta(w_2) \) is non-trivial. Finally, we can also choose Neumann boundary conditions for the gauge field at \( x = L \) and make it dynamical at the boundary. When we apply the sandwich procedure, the boundary theory contains an additional Yang-Mills sector whose 1-form symmetry is given by the field \( c_2 \). It would be interesting to derive the symmetry theory action from a string theory or holographic construction.

\section{SymTh for $\mathbb{Q}/\mathbb{Z}$ 0- and 1-form non-invertible symmetries\label{sec:bulkaxmax}}
We propose a bulk action for the SymTh of non-invertible 0- and 1-form symmetries in 4d, which is given by:
\begin{equation} 
\begin{aligned}\label{eq:SQZ01}
    S_{\rm SymTh} = \int_{M_5} \left( -\frac{1}{2} da_1 \wedge \ast_5 da_1 - \frac{1}{2} dc_1 \wedge \ast_5 dc_1 - \frac{1}{2} db_2 \wedge \ast_5 db_2 - \frac{1}{2} dc_0 \wedge \ast_5 dc_0 \right. 
    \\
    \left. + \frac{1}{2} a_1 \wedge dc_1 \wedge dc_1 + b_2 \wedge dc_1 \wedge dc_0 \right) \, .
    \end{aligned}
\end{equation}
We will derive this action in two ways: one is a partial bottom-up approach directly from axion-Maxwell theory in 4d, and the other is a top-down approach, which involves dimensional reduction of IIB 10d supergravity. Note that there may be an additional term of the form \( a_1 \wedge da_1 \wedge da_1 \), which accounts for the chiral anomaly of the boundary QFT by imposing certain boundary conditions. This term obstructs the choice of Neumann boundary conditions for \( a_1 \), i.e., gauging the \( U(1) \) symmetry whose gauge field is \( a_1 \). For simplicity, we neglect this coupling in the bulk while keeping in mind its impact on boundary conditions. Finally, there seems to be no explicit term corresponding to the mixed anomaly between the electric and magnetic one-form symmetry in the axion-Maxwell theory. However this anomaly is implicitly encoded in the obstruction of choosing certain boundary conditions. For instance from the boundary variational problem, it is not possible to choose Neumann b.c. simultaneously for $c_0,c_1,b_2$. That is an obstruction to gauge both the electric and the magnetic one-form symmetry at the boundary in the axion-Maxwell theory. 

\subsection{Topological couplings from the boundary action}
We propose the action \eqref{eq:SQZ01} as the symmetry theory for 4d axion-Maxwell theory. In particular, we derive the topological couplings in the second line of \eqref{eq:SQZ01} from the 4d axion-Maxwell Lagrangian, which is given by:
\begin{equation}
    S_{\rm aM}^{4d} = - \int_{M_4} \left( \frac{1}{2} d\theta \wedge \ast d\theta + \frac{1}{2} f_2 \wedge \ast f_2 - \frac{K}{2} \theta f_2 \wedge f_2 \right) \, ,
\end{equation}
where \( f_2 = da_1 \), \( \theta \) is a periodic scalar field with \( \theta \sim \theta + 2\pi \), and \( \frac{1}{2} \int_{M_4} f_2 \wedge f_2 \in \mathbb{Z} \) when \( M_4 \) is a closed spin manifold. Additionally, \( K \in \mathbb{Z} \). The theory has four currents:
\begin{equation}
\begin{aligned}
    & J^{(1)}_{s} = d\theta, \\
    & J^{(3)}_{w} = \ast d\theta, \\
    & J^{(2)}_{e} = f_2, \\
    & J^{(2)}_{m} = \ast f_2.
\end{aligned}
\end{equation}
Here, \( s \) stands for shift, \( w \) for winding, \( e \) for electric, and \( m \) for magnetic. The conservation equations for these currents are:
\begin{equation}
\begin{aligned}
    & d\ast J^{(1)}_{s} = -\frac{K}{2} f_2 \wedge f_2, \\
    & d\ast J^{(3)}_{w} = 0, \\
    & d\ast J^{(2)}_{e} = K d\theta \wedge f_2, \\
    & d\ast J^{(2)}_{m} = 0.
\end{aligned}
\end{equation}
From these currents and their conservation equations, we observe that the theory possesses a \( U(1)^{(2)} \) 2-form symmetry and a \( U(1)^{(1)}_m \) magnetic 1-form symmetry. For \( K = 1 \), the theory also has non-invertible 0- and 1-form symmetries \cite{Choi:2022fgx}. These can be derived from the non-conservation equations for \( J^{(1)}_{s} \) and \( J^{(2)}_{e} \), and are characterized by the topological operators:
\begin{equation} \label{eq:topopaxmax}
\begin{aligned}
    & \mathcal{D}_{1/N}^{(0)}(M_3) = \int [Da] \, \exp\left[ 2 \pi i \oint_{M_3} \left( \frac{1}{N} \ast J^{(1)}_{s} - \frac{N}{2} a \wedge da - a \wedge da_1 \right) \right], \\
    & \mathcal{D}_{1/N}^{(1)}(M_2) = \int [D\phi \, Dc] \, \exp\left[ 2 \pi i \oint_{M_2} \left( \frac{1}{N} \ast J^{(2)}_{e} + N \phi \wedge dc + \theta dc + \phi da_1 \right) \right].
\end{aligned}
\end{equation}
Here, \( a \) is a gauge field defined on \( M_3 \), and \( \phi, c \) are fields defined on \( M_2 \). The generalization of these defects for \( p/N \), along with some non-invertible fusion rules, is discussed in \cite{Choi:2022fgx}.

By adding a suitable counterterm, the topological part of the action can be written as
\begin{equation}
    S_{\rm aM}^{4d} \supset -\frac{1}{2} \int_{M_4} \theta f_2 \wedge f_2 + 2 d( \theta c_1 \wedge f_2) = \frac{1}{2} \int_{M_4} \theta f_2 \wedge f_2 + \int_{M_4} d\theta \wedge c_1 \wedge f_2.
\end{equation}

We now turn on background fields for the electric 1-form symmetry \( B^e_2 \) and for the 0-form shift symmetry \( A_1^s \). Even though these are backgrounds for non-invertible symmetries with values in \( \mathbb{Q}/\mathbb{Z} \), we will treat them as \( U(1) \) fields in this context. The transformation parameter of these background fields in 4d axion-Maxwell theory lies in \( \mathbb{Q}/\mathbb{Z} \); however, from the construction of the defects in \cite{Choi:2022fgx}, we have a natural embedding of the symmetry parameter in \( U(1) \). If we were concerned with the anomaly theory of 4d axion-Maxwell theory, we would need to treat the transformation in \( \mathbb{Q}/\mathbb{Z} \). 

To capture the full symmetry theory—and thus the boundary QFTs generated by different boundary conditions of the bulk fields—we extend the symmetry parameter to generically lie in \( U(1) \). We will see that specific boundary conditions will then force the symmetry parameter to lie in \( \mathbb{Q}/\mathbb{Z} \). 

The background fields transform as:
\begin{equation} \label{eq:shift1e}
    B^e_2 \rightarrow B^e_2 + d \Lambda^e_1 \quad \text{and} \quad c_1 \rightarrow c_1 + \Lambda^e_1,
\end{equation}
and
\begin{equation} \label{eq:shift0s}
    A^s_1 \rightarrow A^s_1 + d \Lambda^s_0 \quad \text{and} \quad \theta \rightarrow \theta + \Lambda^s_0.
\end{equation}
These transformations induce ambiguities that are cancelled by the following bulk topological action:
\begin{equation}
    S_{5d}^{\rm top} = \int_{M_5} \frac{1}{2} A_1^s \wedge f_2 \wedge f_2 + d\theta \wedge B_2^e \wedge f_2,
\end{equation}
where \( \partial M_5 = M_4 \). These topological terms are equivalent to those appearing in \eqref{eq:SQZ01} under the identification \( b_2 \leftrightarrow B_2^e \) and \( a_1 \leftrightarrow A_1^s \).

\subsection{SymTh from Type II Supergravity}

The procedure to derive the bulk action for a symmetry theory from 10d supergravity has been outlined in \cite{Apruzzi:2023uma}. This involves considering the flux sector of supergravity dimensionally reduced on a geometric background of the form:
\begin{equation}
    M_{10} = M_{d} \times X_{10-d} \sim M_{d+1} \times L_{10-d-1},
\end{equation}
where the second equation should be understood as rewriting the geometric background in a specific limit (e.g., near-horizon), and \( \partial(X_{10-d}) = L_{10-d-1} \). In \cite{Apruzzi:2023uma}, the focus was on a truncation to the topological sector, whereas for non-finite symmetries, the goal is to retain the kinetic terms in the bulk as well. The starting point action is a formal expression in 11 dimensions, where the extra dimension is auxiliary and provides a democratic treatment of gauge invariance of the supergravity fields:
\begin{equation} \label{eq:IIBsugra}
    S_{10+1} = \int_{M_{10+1}} \left[ F_1 \, dF_9 - F_3 \, dF_7 + \frac{1}{2} F_5 \, dF_5 + H_3 \, dH_7 + H_3 (F_1 F_7 - F_3 F_5) \right].
\end{equation}

We now consider type IIB supergravity on flat spacetimes with the conifold geometry \( C(T^{1,1}) \):
\begin{equation}
    M_{10} = M_4 \times \mathcal{C}(T^{1,1}),
\end{equation}
where the conifold is a real cone over the Sasaki-Einstein space \( T^{1,1} \cong S^2 \times S^3 \). Near the boundary at infinity of \( C(T^{1,1}) \), the full space can be considered as:
\begin{equation}
     M_{10} = M_5 \times L, \qquad L = T^{1,1}.
\end{equation}

We now make an ansatz for the fluxes:
\begin{equation} \label{eq:sugrans}
\begin{aligned}
    F_3 &= d c_0 \wedge \omega_2, \\
    H_3 &= d b_2, \\
    F_5 &= d c_1 \wedge \omega_3 + \ast_5 d c_1 \wedge \omega_2, \\
    F_7 &= \ast_{10} F_3, \\
    H_7 &= \ast_{10} H_3, \\
    F_1 &= F_9 = 0,
\end{aligned}
\end{equation}
where \( \omega_2 \) is the volume form of \( S^2 \) and \( \omega_3 \) is the volume form of \( S^3 \). We also activate the \( U(1) \) isometry Reeb vector \cite{Apruzzi:2022rei, Bah:2023ymy, Cassani:2010na}, \( a_1 \), such that:
\begin{equation} \label{eq:reeb}
    d \omega_2 = 0, \qquad d \omega_3 = -2 \omega_2 \wedge d a_1.
\end{equation}

If we now substitute \eqref{eq:sugrans} and \eqref{eq:reeb} into \eqref{eq:IIBsugra}, and perform an integration over \( L = S^2 \times S^3 \), while also integrating over the extra auxiliary dimension, we obtain the action in \eqref{eq:SQZ01}. This corresponds to the case of the conifold without \( F_5 \) and \( F_3 \) background fluxes. \footnote{The conifold with no flux does not have normalizable modes, and thus it engineers a \( U(1) \) free gauge theory coupled to an axion in 4d. However, this sector is completely decoupled when any other normalizable modes are present. A similar reduction can also be done in IIA, where the roles of \( \omega_2 \) and \( \omega_3 \) are exchanged.}

\subsection{Topological Operators and Their Action}

From the equations of motion of \eqref{eq:SQZ01}, we derive the following (non-)conservation equations:
\begin{equation} \label{eq:nonconsbulkaxmax}
\begin{aligned}
    & d \ast J_2^{a} = d \ast d a_1 = -\frac{1}{2} d c_1 \wedge d c_1, \\
    & d \ast J_2^{c} = d \ast d c_1 = d a_1 \wedge d c_1 + d b_2 \wedge d c_0, \\
    & d \ast J_1^{c} = d \ast d c_0 = d b_2 \wedge d c_1, \\
    & d \ast J_3^{b} = d \ast d b_2 = -d c_1 \wedge d c_0.
\end{aligned}
\end{equation}
This first set of equations leads to four non-invertible symmetries in the bulk, whose topological operators are similar to those of 4d axion-Maxwell theory, as shown in \eqref{eq:topopaxmax}. These operators are:
\begin{equation} \label{eq:noninvdef}
\begin{aligned}
    & \mathcal{D}_{1/N}^{(1,a)}(M_3) = \int [Da] \exp \left[ 2\pi i \oint_{M_3} \left( \frac{1}{N} \ast J_2^a - \frac{N}{2} a \wedge da - a \wedge dc_1 \right) \right], \\
    & \mathcal{D}_{1/N}^{(1,c)}(M_2) = \int [Da D c D \phi D b] \exp \left[ 2\pi i \oint_{M_2} \left( \frac{1}{N} \ast J_2^c + N a \, dc + a \, dc_1 + c \, da_1 - N \phi \, db - \phi \, db_2 - c_0 \, db \right) \right], \\
    & \mathcal{D}_{1/N}^{(0,c)}(M_4) = \int [Da Db] \exp \left[ 2\pi i \oint_{M_4} \left( \frac{1}{N} \ast J_2^c + N a \, db + a \, db_2 + b \, dc_1 \right) \right], \\
    & \mathcal{D}_{1/N}^{(2,c)}(M_2) = \int [D\phi Da] \exp \left[ 2\pi i \oint_{M_2} \left( \frac{1}{N} \ast J_2^c - N \phi \, da - \phi \, dc_1 - c_0 \, da \right) \right].
\end{aligned}
\end{equation}
These operators can be generalized to any transformation with discrete parameter \( p/N \) \cite{Damia:2022bcd, Choi:2022fgx}, where, in general, the transformation parameters lie in \( \mathbb{Q}/\mathbb{Z} \).

From the Bianchi identities, we obtain the following conservation laws:
\begin{equation}
\begin{aligned}
    & d \ast J_3^{a} = d d a_1 = 0, \\
    & d \ast J_3^{c} = d d c_1 = 0, \\
    & d \ast J_4^{c} = d d c_0 = 0, \\
    & d \ast J_2^{b} = d d b_2 = 0.
\end{aligned}
\end{equation}
This leads to the following topological operators:
\begin{equation}
\begin{aligned}
    U^{(2,a)}_{\alpha}(M_2) &= \exp \left(i \alpha \oint_{M_2} \ast J_3^{a} \right), \\
    U^{(2,c)}_{\alpha}(M_2) &= \exp \left(i \alpha \oint_{M_2} \ast J_3^{c} \right), \\
    U^{(3,a)}_{\alpha}(M_1) &= \exp \left(i \alpha \oint_{M_1} \ast J_4^{c} \right), \\
    U^{(3,b)}_{\alpha}(M_3) &= \exp \left(i \alpha \oint_{M_3} \ast J_2^{b} \right),
\end{aligned}
\end{equation}
where \( \alpha \) are periodic parameters in the range \( [0, 2\pi) \) for these \( U(1) \) symmetries. Although each topological operator may have a different \( \alpha \), we use the same symbol for clarity and to avoid overloading the notation.

\subsection{Boundary Conditions}

In this section, we discuss possible boundary conditions for this action. Depending on the boundary condition, we can determine which topological operators are allowed to be fully projected onto the free boundary. The boundary variation of the action reads:
\begin{equation}
\begin{aligned}
    \delta S_{5d} |_{\partial M} &= \int_{\partial M} \left( - \delta a_1 \wedge \ast_5 d a_1 - \delta c_1 \wedge \ast_5 d c_1 - \delta b_2 \wedge \ast_5 d b_2 - \delta c_0 \wedge \ast_5 d c_0 \right) \\
    &\quad + \delta c_1 \wedge a_1 \wedge d c_1 + \delta c_1 \wedge b_2 \wedge d c_0 + \delta c_0 \, d c_1 \wedge b_2.
\end{aligned}
\end{equation}
We recall that by adding the Chern-Simons counterterm, we do not change the theory, but we can shift the variation of one field to another. Therefore, it is enough to have a field with Dirichlet boundary conditions in the Chern-Simons coupling at the boundary for that term to vanish. There are many possible boundary conditions, each leading to different boundary QFTs when we implement the sandwich procedure. Here, we focus on a single case: having an interacting QFT at the \( x = 0 \) boundary, with the following boundary conditions at \( x = L \):
\begin{equation}
    \delta a_1 |_{x = L} = 0, \qquad \ast_5 d c_1 |_{x = L} = d \ast_4 d c_1, \qquad \delta b_2 = 0, \qquad \ast_5 d c_0 |_{x = L} = d \ast_4 d c_0.
\end{equation}
The second and fourth conditions correspond to the modified Neumann boundary conditions that account for the singleton sector at \( x = L \) and the dynamical gauging of the \( c_1 \) and \( c_0 \) fields. This follows the analysis in \ref{sec:maxw}. The topological operators that are allowed to end on the boundary are those that link the Wilson surfaces of the fields \( a_1 \) and \( b_2 \), which have Dirichlet boundary conditions. These operators are \( \mathcal{D}_{1/N}^{(1,a)}(M_3) \) and \( \mathcal{D}_{1/N}^{(2,c)}(M_2) \). Following the analysis in section \ref{sec:maxw}, when we apply the sandwich and interval compactification procedure, we obtain a \( d \)-dimensional QFT where the operators generate a 0-form and 1-form non-invertible symmetry with transformation parameters labeled by \( p/N \). Moreover, the Neumann boundary conditions for the fields \( c_1 \) and \( c_0 \) indicate that the dual surfaces can end on the boundary, and the operators \( U^{(2,c)}_{\alpha}(M_2) \) and \( U^{(3,a)}_{\alpha}(M_1) \) will then be projected parallel to the boundary at \( x = L \). Thus, they generate continuous \( U(1) \) invertible 1- and 2-form symmetries for the boundary QFT when we compactify the interval. The resulting QFT enjoys these symmetries, and the theory described is indeed 4d axion-Maxwell theory, for which we have described its SymTh. The remaining topological operators link closed Wilson surfaces in the bulk and cannot be projected onto the boundary. From the perspective of our sandwich and interval compactification procedure, they do not act faithfully on the spectrum of boundary operators.

\section{Topological Defects as Branes\label{sec:branes}}
When a QFT is constructed holographically or via geometric engineering, the topological defects for finite symmetries are described by branes \cite{Apruzzi:2022rei,GarciaEtxebarria:2022vzq,Heckman:2022muc,Heckman:2022xgu, Etheredge:2023ler, Bah:2023ymy} in a specific decoupling limit \cite{Apruzzi:2023uma, Heckman:2024oot, Apruzzi:2022rei, Witten:1998xy}. Branes provide an ultraviolet description of the topological defects. In a well-defined topological truncation, they are very useful in describing the charges and their relation to the symmetry TFT \cite{Apruzzi:2023uma}. The question we aim to address here is whether we can argue from an infrared point of view that the defects should be realized by branes. Finally, we will see how this is useful for describing the quantum Hall states that dress the non-invertible $\mathbb{Q}/\mathbb{Z}$ defects of the bulk SymTh studied in section \ref{sec:bulkaxmax}.

\subsection{Ultraviolet Point of View, Approximate Symmetries, and Branes}
We will argue for the presence of the branes in the UV by assuming that the symmetries of the SymTh, when consistently coupled to gravity, must be at most approximate at low energies, following the original hypothesis in \cite{Banks:2010zn, Polchinski:2003bq} that there are no global symmetries in quantum gravity.

\subsubsection*{BF-Theory}
Let us first look at the BF theory describing finite global p-form symmetries:
\begin{equation}
    S_{\rm BF} = N \int_{M_{d+1}} a_{p+1} \wedge d b_{d-p-1}
\end{equation}
where $a_{p+1}$ and $b_{d-p-1}$ are originally $U(1)$ gauge fields but are constrained to be flat by the equations of motion:
\begin{equation}
    N d a_{p+1} = N d b_{d-p-1} = 0
\end{equation}
This implies that we have $\mathbb{Z}_N$-valued holonomies:
\begin{equation}
    N \oint_{\Sigma_{p+1}} a_{p+1} \in \mathbb{Z}, \qquad N \oint_{\Sigma_{d-p-1}} b_{d-p-1} \in \mathbb{Z}.
\end{equation}
The space $M_{d+1}$ has a boundary, $\partial M_{d+1} = M_d$. The topological operators are \cite{Brennan:2023mmt}:
\begin{equation}
    U_{\frac{k}{N}}(\Sigma_{p+1}) = e^{2 \pi i k \oint_{\Sigma_{p+1}} a_{p+1}}, \qquad U_{\frac{k}{N}}(\Sigma_{d-p-1}) = e^{2 \pi i k \oint_{\Sigma_{d-p-1}} b_{d-p-1}}
\end{equation}
These generate the symmetries $\mathbb{Z}_N^{(d-p-1)}$ and $\mathbb{Z}_N^{(p+1)}$, respectively. We can ask what the UV fate of these symmetries is when we couple the theory to gravity. 
Let us start from $\mathbb{Z}^{(p+1)}_N$.
If this is not a 0-form symmetry, which is usually approximate in the IR and violated by irrelevant operators, there must be a dynamical $p$-dimensional (including the time direction) object that terminates the $U_{\frac{k}{N}}(\Sigma_{p+1})$ operators (screening) \cite{Cordova:2022rer}, making the $U_{\frac{k}{N}}(\Sigma_{d-p-1})$ non topological anymore. A good candidates for these dynamical objects are branes. In this case, we need a $(p)$-brane, which is responsible for making this symmetry approximate below its tension scale and is the one charged under $a_{p+1}$. For the $\mathbb{Z}^{d-p-1}_N$ the roles of $ U_{\frac{k}{N}}(\Sigma_{p+1})$ and $ U_{\frac{k}{N}}(\Sigma_{d-p-1})$ are exchanged. Under the assumption of the validity of the no-global-symmetry hypothesis, the brane that must be present in the UV is the one charged under $b_{d-p-1}$, i.e., a $(d-p-2)$-brane. Therefore, by assuming the validity of the hypothesis that the symmetries of the bulk theory must be approximate in the IR when consistently coupled to gravity, we provide evidence for the presence of the branes charged under $a_{p+1}$ and $b_{d-p-1}$. These are good UV condidates for the topological defects, that become symmetry defects in the topological limit/truncation defined in \cite{Apruzzi:2023uma}, i.e., when these branes are taken to the boundary.

\subsubsection*{Maxwell}
We can run similar arguments for the $(p+1)$-form Maxwell theory. We have analyzed this case in detail in \ref{sec:bulkaxmax}. The topological operators are given by \eqref{eq:topopmax}, and the Wilson surface operators are in \eqref{eq:WSmax}. One of the two symmetries will be broken by the Dirichlet boundary condition, making the corresponding Wilson surface end on the boundary. The other one will instead be only approximate in the IR when coupling Maxwell to gravity. Suppose we choose Dirichlet boundary conditions for $b_{d-p-2}$. This implies that $V_m(M_{d-p-2})$ can end on the boundary, and the symmetry generated by $U_{\beta}(\Sigma_{p+2})$ is automatically broken, while the symmetry generated by $U_{\alpha}(\Sigma_{d-p-2})$ is only approximate. This occurs by postulating that $W_q(M_{p+1})$ are screened by the dynamical $p$-brane present in the UV theory. At the scale of the brane tension, this can be seen from the source equation for the equation of motion, which breaks the conservation equation for $J_{p+1}$:
\begin{equation}
    d \ast_{d+1} F_{p+1} = \ast_{d+1} j_{p+1}
\end{equation}
where $\ast_{d+1} j_{p+1}$ is the magnetic source of a $p$-brane.

A similar conclusion holds when choosing Dirichlet boundary conditions for $a_{p+1}$, which implies the existence of $(d-p-3)$-branes. This can be seen from the source equation for the Bianchi identity:
\begin{equation}
    d F_{p+1} = \ast_{d+1} j_{d-p-2}
\end{equation}
where $\ast_{d+1} j_{d-p-2}$ is the magnetic source of a $(d-p-3)$-brane. So far, these dynamical objects do not provide avatars for any topological operators. 

We will see in the next subsection that the predicted branes can still give the quantum Hall states necessary to define non-invertible topological operators \footnote{A different perspective is given in \cite{Cvetic:2023plv}, where the continuous symmetries are described by flux branes.}. This will occur when we include mixed Chern-Simons couplings beyond the free Maxwell kinetic terms.

\subsection{Branes in the SymTh of 4d Axion-Maxwell}
Let us now consider the bulk theory \eqref{eq:SQZ01}. We can make the symmetries approximate below the brane tension scales by adding various brane sources for the conservation equation. In addition, since we know the string theory origin from the reduction of type IIB on $T^{1,1} = S^2 \times S^3$ \eqref{eq:sugrans}, we can identify the branes responsible for this. For instance, from the IIB supergravity Bianchi identities with brane sources, we have:
\begin{equation} \label{eq:Bws}
\begin{aligned}
    & d \ast J_3^{c} = d dc_1 = \ast_{5} j_{2} (\text{D3 on } S^2) = \delta^{(3)}_{\text{D3}(S^2)} \\
    & d \ast J_4^{c} = d dc_0 = \ast_{5} j_{3} (\text{D5 on } S^3) = \delta^{(2)}_{\text{D5}(S^3)} \\
    & d \ast J_2^{b} = d db_2 = \ast_{5} j_{1} (\text{NS5 on } S^2 \times S^3) = \delta^{(4)}_{\text{NS}(S^3 \times S^2)}
\end{aligned}
\end{equation}
We can now examine the non-conservation equations \eqref{eq:nonconsbulkaxmax} and what these sources imply for them. In particular, \eqref{eq:nonconsbulkaxmax} will not only define the non-invertible defects in \eqref{eq:noninvdef}, but, in the absence of sources, it also defines Chern-Weil conserved currents. This is simply shown by applying an extra derivative operator on \eqref{eq:nonconsbulkaxmax}. Let us focus on the following two equations of motion:
\begin{equation}
\begin{aligned}
    & 0 = dd \ast da_1 = \frac{1}{2} d (dc_1 \wedge dc_1) = d \ast_{5} J_1^{\rm CW} \\
    & 0 = dd \ast db_2 = d (dc_1 \wedge dc_0) = d \ast_{5} J_2^{\rm CW}
\end{aligned}
\end{equation}
If we now allow for sources like those in \eqref{eq:Bws}, we get inconsistencies with the equations of motion:
\begin{equation}
\begin{aligned}
    & 0 = dd \ast da_1 = \frac{1}{2} d (dc_1 \wedge dc_1) = d \ast_{5} J_1^{\rm CW} = \delta^{(3)}_{\rm D3} \wedge dc_1 \neq 0 \\
    & 0 = dd \ast db_2 = d (dc_1 \wedge dc_0) = d \ast_{5} J_2^{\rm CW} = \delta^{(3)}_{\rm D3} \wedge dc_0 + dc_1 \wedge \delta_{\rm D5}^{(2)} \neq 0
\end{aligned}
\end{equation}
This can be fixed by adding worldvolume gauge fields on the brane \cite{Heidenreich:2021xpr}:
\begin{equation}
\begin{aligned}
    & \frac{1}{2} d (dc_1 \wedge dc_1 - 2 f^{\rm D3}_1 \wedge \delta^{(3)}_{\rm D3}) = 0 \\
    & d (dc_1 \wedge dc_0 - f^{\rm D3}_0 \wedge \delta^{(3)}_{\rm D3} - f^{\rm D5}_1 \wedge \delta_{\rm D5}^{(2)}) = 0
\end{aligned}
\end{equation}
where $f^{\rm D3}_1$, $f^{\rm D5}_1$, and $f^{\rm D3}_0$ are the field strengths living on the brane worldvolume. This is related to the usual $a$ gauge field by Hodge duality in the space that the brane spans and wraps inside $M_{d+1} \times L$. In addition, they must satisfy:
\begin{equation}
    df^{\rm D3}_1 = dc_1, \qquad df^{\rm D3}_0 = dc_0, \qquad df^{\rm D5}_1 = dc_1
\end{equation}
This leads to St\"{u}ckelberg couplings on the brane worldvolume and describes how the gauge field on the brane couples to the bulk fields. For instance, we can dualize the $f^{\rm D3}_1$, $f^{\rm D5}_1$, and $f^{\rm D3}_0$ appearing in the St\"{u}ckelberg terms to obtain BF terms for the brane worldvolume \cite{Maldacena:2001ss}:
\begin{equation}
\begin{aligned}
    & S_{\text{D3}(S^2)} \subset \int_{\Sigma_2} a \wedge dc_0 + \phi dc_1 \\
    & S_{\text{D5}(S^3)} \subset \int_{\Sigma_3} \tilde a \wedge dc_1
\end{aligned}
\end{equation}
where $a$ is the dual field of $f^{\rm D3}_0$ in $\Sigma_2$, $\phi$ is the dual of $f^{\rm D3}_1$ in $\Sigma_2$, and $\tilde a$ is the dual of $f^{\rm D5}_1$ in $\Sigma_3$. A full analysis of the branes and their worldvolume gauge fields is beyond the scope of this paper, and we plan to revisit it in the future. Nevertheless, this already provides strong evidence that, in the topological limit defined in \cite{Apruzzi:2023uma}, the D3 brane on $S^2$ produces the topological dressing of the $\mathcal{D}_{1/N}^{(1,a)}(M_3)$ defect, and the D5 brane on $S^3$ provides the topological dressing of the $\mathcal{D}_{1/N}^{(2,c)}(M_2)$ non-invertible defect. It will be interesting to study the worldvolume theories of these branes in a more systematic fashion by reducing the DBI and Wess-Zumino effective action and taking the topological limit.

\section{Conclusions and Outlook}

In this work, we explored the possibility of using $(p+1)$-form Maxwell theory (free in some effective regimes depending on the dimension) as a SymTh on the product manifold $I_L \times M_d$, where a $\text{QFT}_d$ lives on $M_d$ with a continuous $U(1)^{(p)}$ symmetry. We also considered cases where the Maxwell theory is decorated by Chern-Simons couplings, leading to 2-groups or non-invertible $\mathbb{Q}/\mathbb{Z}$ symmetries. We classified the possible boundary conditions that lead to a consistent "sandwich" procedure, which avoids strong coupling regimes—a feature that must be taken into account when the bulk theories are not topological. Our proposal has been applied to various examples, ranging from two-dimensional gauge theories to higher-dimensional models with 2-group structures and non-invertible symmetries. 

In the final part of the paper, we also derived the SymTh for non-finite, non-invertible symmetries from IIB supergravity, closely related to the boundary 4d axion-Maxwell theory. Additionally, we provided evidence for the UV origin of the quantum Hall states in terms of string theory branes wrapping $S^2$ or $S^3$ of the conifold geometry. 

There are several exciting directions for future research and questions to address. As a further low-dimensional example, it would be interesting to investigate applications to $\text{Maxwell}_3$ and $\text{YM}_3$, which are not as nearly topological as their two-dimensional versions, when coupled to a 2d boundary QFT. Of particular interest in this context (but also more generally) would be to explore the consequences of coupling the bulk theory to a conformal theory on the boundary. Specifically, it would be valuable to investigate whether the bulk theory is able to capture the symmetry structure of the conformal group. 

Additionally, it would be beneficial to systematically analyze all the boundary conditions and topological operators of the SymTh for the 4d axion-Maxwell theory and generalize the results to other dimensions. Another key aspect to address is the full analysis of the worldvolume theory of the branes to match the complete Lagrangian of the quantum Hall states of the non-invertible topological defects. Furthermore, it would be intriguing to examine what the topological operators in the bulk would yield if we make them end on the boundary rather than project them parallel to it. Finally, understanding the condensation defects of these topological defects would bring us closer to a categorical formulation of non-finite symmetries. In order to fully build the SymTh framework it is necessary to study also interfaces that are not necessarily topological. This will allow to include electromagnetic dual symmetries of the boundary theory as well as possible symmetry (non-)topological manipulations.

\section*{Acknowledgement}
We thank Sakura Sch\"afer-Nameki, Noppadol Mekareeya, Andrea Antinucci, Luca Martucci, Luigi Tizzano, Ingo Runkel, Cristian Copetti, Max H\"ubner for discussions. 
The work of FA is supported in part by the Italian MUR Departments of Excellence grant 2023-2027 "Quantum Frontiers”. ND acknowledges the receipt of the joint grant from the Abdus Salam International Centre for Theoretical Physics (ICTP), Trieste, Italy and INFN, sede centrale Frascati, Italy.

\newpage

\bibliography{Symbib}

@article{Migdal:1975zg,
    author = "Migdal, Alexander A.",
    editor = "Khalatnikov, I. M. and Mineev, V. P.",
    title = "{Recursion equations in gauge field theories}",
    reportNumber = "PRINT-75-1043 (LANDAU-INST)",
    journal = "Sov. Phys. JETP",
    volume = "42",
    pages = "413--418",
    year = "1975"
}

@article{Rusakov90,
author = {RUSAKOV, B. YE.},
title = {LOOP AVERAGES AND PARTITION FUNCTIONS IN U(N) GAUGE THEORY ON TWO-DIMENSIONAL MANIFOLDS},
journal = {Modern Physics Letters A},
volume = {05},
number = {09},
pages = {693-703},
year = {1990},
doi = {10.1142/S0217732390000780},
}

@article{blau1992quantum,
  title={Quantum Yang-Mills theory on arbitrary surfaces},
  author={Blau, Matthias and Thompson, George},
  journal={International Journal of Modern Physics A},
  volume={7},
  number={16},
  pages={3781--3806},
  year={1992},
  publisher={World Scientific}
}

@article{Lake:2018dqm,
    author = "Lake, Ethan",
    title = "{Higher-form symmetries and spontaneous symmetry breaking}",
    eprint = "1802.07747",
    archivePrefix = "arXiv",
    primaryClass = "hep-th",
    month = "2",
    year = "2018"
}

@article{Pulmann:2019vrw,
    author = "Pulmann, J\'an and \v{S}evera, Pavol and Valach, Fridrich",
    title = "{A nonabelian duality for (higher) gauge theories}",
    eprint = "1909.06151",
    archivePrefix = "arXiv",
    primaryClass = "hep-th",
    doi = "10.4310/ATMP.2021.v25.n1.a5",
    journal = "Adv. Theor. Math. Phys.",
    volume = "25",
    number = "1",
    pages = "241--274",
    year = "2021"
}

@article{Witten:1992xu,
    author = "Witten, Edward",
    title = "{Two-dimensional gauge theories revisited}",
    eprint = "hep-th/9204083",
    archivePrefix = "arXiv",
    doi = "10.1016/0393-0440(92)90034-X",
    journal = "J. Geom. Phys.",
    volume = "9",
    pages = "303--368",
    year = "1992"
}

@article{Witten:1991we,
    author = "Witten, Edward",
    title = "{On quantum gauge theories in two-dimensions}",
    doi = "10.1007/BF02100009",
    journal = "Commun. Math. Phys.",
    volume = "141",
    pages = "153--209",
    year = "1991"
}

@article{Blau:1991mp,
    author = "Blau, Matthias and Thompson, George",
    title = "{Quantum Yang-Mills theory on arbitrary surfaces}",
    reportNumber = "NIKHEF-H-91-09, MZ-TH-91-17",
    doi = "10.1142/S0217751X9200168X",
    journal = "Int. J. Mod. Phys. A",
    volume = "7",
    pages = "3781--3806",
    year = "1992"
}

@article{Nguyen:2021naa,
    author = {Nguyen, Mendel and Tanizaki, Yuya and \"Unsal, Mithat},
    title = "{Noninvertible 1-form symmetry and Casimir scaling in 2D Yang-Mills theory}",
    eprint = "2104.01824",
    archivePrefix = "arXiv",
    primaryClass = "hep-th",
    reportNumber = "YITP-21-29",
    doi = "10.1103/PhysRevD.104.065003",
    journal = "Phys. Rev. D",
    volume = "104",
    number = "6",
    pages = "065003",
    year = "2021"
}

@inbook{Gukov:2014gja,
    author = "Gukov, Sergei",
    editor = {Teschner, J\"org},
    title = "{Surface Operators}",
    booktitle = "{New Dualities of Supersymmetric Gauge Theories}",
    eprint = "1412.7127",
    archivePrefix = "arXiv",
    primaryClass = "hep-th",
    doi = "10.1007/978-3-319-18769-3_8",
    pages = "223--259",
    year = "2016"
}

@article{Gukov:2008sn,
    author = "Gukov, Sergei and Witten, Edward",
    title = "{Rigid Surface Operators}",
    eprint = "0804.1561",
    archivePrefix = "arXiv",
    primaryClass = "hep-th",
    doi = "10.4310/ATMP.2010.v14.n1.a3",
    journal = "Adv. Theor. Math. Phys.",
    volume = "14",
    number = "1",
    pages = "87--178",
    year = "2010"
}

@article{Apruzzi:2023uma,
    author = "Apruzzi, Fabio and Bonetti, Federico and Gould, Dewi S. W. and Schafer-Nameki, Sakura",
    title = "{Aspects of Categorical Symmetries from Branes: SymTFTs and Generalized Charges}",
    eprint = "2306.16405",
    archivePrefix = "arXiv",
    primaryClass = "hep-th",
    month = "6",
    year = "2023"
}

@article{Bah:2023ymy,
    author = "Bah, Ibrahima and Leung, Enoch and Waddleton, Thomas",
    title = "{Non-Invertible Symmetries, Brane Dynamics, and Tachyon Condensation}",
    eprint = "2306.15783",
    archivePrefix = "arXiv",
    primaryClass = "hep-th",
    month = "6",
    year = "2023"
}

@article{Hofman:2017vwr,
    author = "Hofman, Diego M. and Iqbal, Nabil",
    title = "{Generalized global symmetries and holography}",
    eprint = "1707.08577",
    archivePrefix = "arXiv",
    primaryClass = "hep-th",
    doi = "10.21468/SciPostPhys.4.1.005",
    journal = "SciPost Phys.",
    volume = "4",
    number = "1",
    pages = "005",
    year = "2018"
}

@article{Heckman:2024oot,
    author = {Heckman, Jonathan J. and H\"ubner, Max and Murdia, Chitraang},
    title = "{On the Holographic Dual of a Topological Symmetry Operator}",
    eprint = "2401.09538",
    archivePrefix = "arXiv",
    primaryClass = "hep-th",
    month = "1",
    year = "2024"
}

@article{Banks:2010zn,
    author = "Banks, Tom and Seiberg, Nathan",
    title = "{Symmetries and Strings in Field Theory and Gravity}",
    eprint = "1011.5120",
    archivePrefix = "arXiv",
    primaryClass = "hep-th",
    doi = "10.1103/PhysRevD.83.084019",
    journal = "Phys. Rev. D",
    volume = "83",
    pages = "084019",
    year = "2011"
}

@article{Seiberg:2018ntt,
    author = "Seiberg, Nathan and Tachikawa, Yuji and Yonekura, Kazuya",
    title = "{Anomalies of Duality Groups and Extended Conformal Manifolds}",
    eprint = "1803.07366",
    archivePrefix = "arXiv",
    primaryClass = "hep-th",
    reportNumber = "IPMU-18-0044",
    doi = "10.1093/ptep/pty069",
    journal = "PTEP",
    volume = "2018",
    number = "7",
    pages = "073B04",
    year = "2018"
}

@article{Horowitz:1989ng,
    author = "Horowitz, Gary T.",
    title = "{Exactly Soluble Diffeomorphism Invariant Theories}",
    reportNumber = "NSF-ITP-88-178",
    doi = "10.1007/BF01218410",
    journal = "Commun. Math. Phys.",
    volume = "125",
    pages = "417",
    year = "1989"
}

@article{Kravec:2013pua,
    author = "Kravec, S. M. and McGreevy, John",
    title = "{A gauge theory generalization of the fermion-doubling theorem}",
    eprint = "1306.3992",
    archivePrefix = "arXiv",
    primaryClass = "hep-th",
    reportNumber = "UCSD-PTH-13-07, MIT-CTP-4468",
    doi = "10.1103/PhysRevLett.111.161603",
    journal = "Phys. Rev. Lett.",
    volume = "111",
    pages = "161603",
    year = "2013"
}

@article{Kravec:2014aza,
    author = "Kravec, S. M. and McGreevy, John and Swingle, Brian",
    title = "{All-fermion electrodynamics and fermion number anomaly inflow}",
    eprint = "1409.8339",
    archivePrefix = "arXiv",
    primaryClass = "hep-th",
    reportNumber = "UCSD-PTH-14-06",
    doi = "10.1103/PhysRevD.92.085024",
    journal = "Phys. Rev. D",
    volume = "92",
    number = "8",
    pages = "085024",
    year = "2015"
}

@article{Aharony:1998qu,
    author = "Aharony, Ofer and Witten, Edward",
    title = "{Anti-de Sitter space and the center of the gauge group}",
    eprint = "hep-th/9807205",
    archivePrefix = "arXiv",
    reportNumber = "IASSNS-HEP-98-66, RU-98-34",
    doi = "10.1088/1126-6708/1998/11/018",
    journal = "JHEP",
    volume = "11",
    pages = "018",
    year = "1998"
}

@article{Shao:2023gho,
    author = "Shao, Shu-Heng",
    title = "{What's Done Cannot Be Undone: TASI Lectures on Non-Invertible Symmetry}",
    eprint = "2308.00747",
    archivePrefix = "arXiv",
    primaryClass = "hep-th",
    reportNumber = "YITP-SB-2023-19",
    month = "8",
    year = "2023"
}

@article{Putrov:2022pua,
    author = "Putrov, Pavel",
    title = "{$\mathbb{Q}/\mathbb{Z}$ symmetry}",
    eprint = "2208.12071",
    archivePrefix = "arXiv",
    primaryClass = "hep-th",
    month = "8",
    year = "2022"
}

@article{Bonetti:2024cjk,
    author = "Bonetti, Federico and Del Zotto, Michele and Minasian, Ruben",
    title = "{SymTFTs for Continuous non-Abelian Symmetries}",
    eprint = "2402.12347",
    archivePrefix = "arXiv",
    primaryClass = "hep-th",
    month = "2",
    year = "2024"
}

@article{Bhardwaj:2023fca,
    author = "Bhardwaj, Lakshya and Bottini, Lea E. and Pajer, Daniel and Schafer-Nameki, Sakura",
    title = "{Categorical Landau Paradigm for Gapped Phases}",
    eprint = "2310.03786",
    archivePrefix = "arXiv",
    primaryClass = "cond-mat.str-el",
    month = "10",
    year = "2023"
}

@article{Antinucci:2024zjp,
    author = "Antinucci, Andrea and Benini, Francesco",
    title = "{Anomalies and gauging of U(1) symmetries}",
    eprint = "2401.10165",
    archivePrefix = "arXiv",
    primaryClass = "hep-th",
    reportNumber = "SISSA 01/2024/FISI",
    month = "1",
    year = "2024"
}

@article{Brennan:2024fgj,
    author = "Brennan, T. Daniel and Sun, Zhengdi",
    title = "{A SymTFT for Continuous Symmetries}",
    eprint = "2401.06128",
    archivePrefix = "arXiv",
    primaryClass = "hep-th",
    month = "1",
    year = "2024"
}

@article{Gaiotto:2017yup,
    author = "Gaiotto, Davide and Kapustin, Anton and Komargodski, Zohar and Seiberg, Nathan",
    title = "{Theta, Time Reversal, and Temperature}",
    eprint = "1703.00501",
    archivePrefix = "arXiv",
    primaryClass = "hep-th",
    doi = "10.1007/JHEP05(2017)091",
    journal = "JHEP",
    volume = "05",
    pages = "091",
    year = "2017"
}

@article{Carta:2022fxc,
    author = "Carta, Federico and Giacomelli, Simone and Mekareeya, Noppadol and Mininno, Alessandro",
    title = "{A tale of 2-groups: D$_{p}$(USp(2N)) theories}",
    eprint = "2208.11130",
    archivePrefix = "arXiv",
    primaryClass = "hep-th",
    reportNumber = "ZMP-HH/22-16",
    doi = "10.1007/JHEP06(2023)102",
    journal = "JHEP",
    volume = "06",
    pages = "102",
    year = "2023"
}

@article{Mekareeya:2022spm,
    author = "Mekareeya, Noppadol and Sacchi, Matteo",
    title = "{Mixed anomalies, two-groups, non-invertible symmetries, and 3d superconformal indices}",
    eprint = "2210.02466",
    archivePrefix = "arXiv",
    primaryClass = "hep-th",
    doi = "10.1007/JHEP01(2023)115",
    journal = "JHEP",
    volume = "01",
    pages = "115",
    year = "2023"
}

@article{Polchinski:2003bq,
    author = "Polchinski, Joseph",
    editor = "Baer, H. and Belyaev, A.",
    title = "{Monopoles, duality, and string theory}",
    eprint = "hep-th/0304042",
    archivePrefix = "arXiv",
    doi = "10.1142/S0217751X0401866X",
    journal = "Int. J. Mod. Phys. A",
    volume = "19S1",
    pages = "145--156",
    year = "2004"
}

@article{Cassani:2010na,
    author = "Cassani, Davide and Faedo, Anton F.",
    title = "{A Supersymmetric consistent truncation for conifold solutions}",
    eprint = "1008.0883",
    archivePrefix = "arXiv",
    primaryClass = "hep-th",
    reportNumber = "DFPD-2010-TH-14",
    doi = "10.1016/j.nuclphysb.2010.10.010",
    journal = "Nucl. Phys. B",
    volume = "843",
    pages = "455--484",
    year = "2011"
}

@article{DiPietro:2019hqe,
    author = "Di Pietro, Lorenzo and Gaiotto, Davide and Lauria, Edoardo and Wu, Jingxiang",
    title = "{3d Abelian Gauge Theories at the Boundary}",
    eprint = "1902.09567",
    archivePrefix = "arXiv",
    primaryClass = "hep-th",
    doi = "10.1007/JHEP05(2019)091",
    journal = "JHEP",
    volume = "05",
    pages = "091",
    year = "2019"
}

@article{Damia:2022seq,
    author = "Damia, Jeremias Aguilera and Argurio, Riccardo and Tizzano, Luigi",
    title = "{Continuous Generalized Symmetries in Three Dimensions}",
    eprint = "2206.14093",
    archivePrefix = "arXiv",
    primaryClass = "hep-th",
    doi = "10.1007/JHEP05(2023)164",
    journal = "JHEP",
    volume = "23",
    pages = "164",
    year = "2023"
}

@article{Antinucci:2022cdi,
    author = "Antinucci, Andrea and Copetti, Christian and Galati, Giovanni and Rizi, Giovanni",
    title = "{''Zoology'' of non-invertible duality defects: the view from class $\mathcal{S}$}",
    eprint = "2212.09549",
    archivePrefix = "arXiv",
    primaryClass = "hep-th",
    month = "12",
    year = "2022"
}

@article{Schafer-Nameki:2023jdn,
    author = "Schafer-Nameki, Sakura",
    title = "{ICTP Lectures on (Non-)Invertible Generalized Symmetries}",
    eprint = "2305.18296",
    archivePrefix = "arXiv",
    primaryClass = "hep-th",
    month = "5",
    year = "2023"
}

@article{Brennan:2023mmt,
    author = "Brennan, T. Daniel and Hong, Sungwoo",
    title = "{Introduction to Generalized Global Symmetries in QFT and Particle Physics}",
    eprint = "2306.00912",
    archivePrefix = "arXiv",
    primaryClass = "hep-ph",
    month = "6",
    year = "2023"
}

@article{Witten:1998xy,
    author = "Witten, Edward",
    title = "{Baryons and branes in anti-de Sitter space}",
    eprint = "hep-th/9805112",
    archivePrefix = "arXiv",
    reportNumber = "IASSNS-HEP-98-42",
    doi = "10.1088/1126-6708/1998/07/006",
    journal = "JHEP",
    volume = "07",
    pages = "006",
    year = "1998"
}

@article{Maldacena:2001ss,
    author = "Maldacena, Juan Martin and Moore, Gregory W. and Seiberg, Nathan",
    title = "{D-brane charges in five-brane backgrounds}",
    eprint = "hep-th/0108152",
    archivePrefix = "arXiv",
    reportNumber = "RUNHETC-2001-25",
    doi = "10.1088/1126-6708/2001/10/005",
    journal = "JHEP",
    volume = "10",
    pages = "005",
    year = "2001"
}

@article{Belov:2004ht,
    author = "Belov, Dmitriy and Moore, Gregory W.",
    title = "{Conformal blocks for AdS(5) singletons}",
    eprint = "hep-th/0412167",
    archivePrefix = "arXiv",
    month = "12",
    year = "2004"
}

@article{Bartsch:2023wvv,
    author = "Bartsch, Thomas and Bullimore, Mathew and Grigoletto, Andrea",
    title = "{Representation theory for categorical symmetries}",
    eprint = "2305.17165",
    archivePrefix = "arXiv",
    primaryClass = "hep-th",
    month = "5",
    year = "2023"
}

@article{Bhardwaj:2023ayw,
    author = "Bhardwaj, Lakshya and Schafer-Nameki, Sakura",
    title = "{Generalized Charges, Part II: Non-Invertible Symmetries and the Symmetry TFT}",
    eprint = "2305.17159",
    archivePrefix = "arXiv",
    primaryClass = "hep-th",
    month = "5",
    year = "2023"
}

@article{Bhardwaj:2022dyt,
    author = "Bhardwaj, Lakshya and Bullimore, Mathew and Ferrari, Andrea E. V. and Schafer-Nameki, Sakura",
    title = "{Anomalies of Generalized Symmetries from Solitonic Defects}",
    eprint = "2205.15330",
    archivePrefix = "arXiv",
    primaryClass = "hep-th",
    month = "5",
    year = "2022"
}

@article{Benini:2018reh,
    author = "Benini, Francesco and C\'ordova, Clay and Hsin, Po-Shen",
    title = "{On 2-Group Global Symmetries and their Anomalies}",
    eprint = "1803.09336",
    archivePrefix = "arXiv",
    primaryClass = "hep-th",
    reportNumber = "SISSA 10/2018/FISI, SISSA-10-2018-FISI",
    doi = "10.1007/JHEP03(2019)118",
    journal = "JHEP",
    volume = "03",
    pages = "118",
    year = "2019"
}

@article{Gaiotto:2014kfa,
    author = "Gaiotto, Davide and Kapustin, Anton and Seiberg, Nathan and Willett, Brian",
    title = "{Generalized Global Symmetries}",
    eprint = "1412.5148",
    archivePrefix = "arXiv",
    primaryClass = "hep-th",
    doi = "10.1007/JHEP02(2015)172",
    journal = "JHEP",
    volume = "02",
    pages = "172",
    year = "2015"
}

@article{Apruzzi:2021nmk,
    author = "Apruzzi, Fabio and Bonetti, Federico and Etxebarria, I\~naki Garc\'\i{}a and Hosseini, Saghar S. and Schafer-Nameki, Sakura",
    title = "{Symmetry TFTs from String Theory}",
    eprint = "2112.02092",
    archivePrefix = "arXiv",
    primaryClass = "hep-th",
    month = "12",
    year = "2021"
}

@article{Freed:2022qnc,
    author = "Freed, Daniel S. and Moore, Gregory W. and Teleman, Constantin",
    title = "{Topological symmetry in quantum field theory}",
    eprint = "2209.07471",
    archivePrefix = "arXiv",
    primaryClass = "hep-th",
    month = "9",
    year = "2022"
}

@article{Heidenreich:2021xpr,
    author = "Heidenreich, Ben and McNamara, Jacob and Montero, Miguel and Reece, Matthew and Rudelius, Tom and Valenzuela, Irene",
    title = "{Non-invertible global symmetries and completeness of the spectrum}",
    eprint = "2104.07036",
    archivePrefix = "arXiv",
    primaryClass = "hep-th",
    reportNumber = "ACFI-T21-03",
    doi = "10.1007/JHEP09(2021)203",
    journal = "JHEP",
    volume = "09",
    pages = "203",
    year = "2021"
}

@article{Gaiotto:2020iye,
    author = "Gaiotto, Davide and Kulp, Justin",
    title = "{Orbifold groupoids}",
    eprint = "2008.05960",
    archivePrefix = "arXiv",
    primaryClass = "hep-th",
    doi = "10.1007/JHEP02(2021)132",
    journal = "JHEP",
    volume = "02",
    pages = "132",
    year = "2021"
}

@article{Cordova:2022ieu,
    author = "Cordova, Clay and Ohmori, Kantaro",
    title = "{Noninvertible Chiral Symmetry and Exponential Hierarchies}",
    eprint = "2205.06243",
    archivePrefix = "arXiv",
    primaryClass = "hep-th",
    doi = "10.1103/PhysRevX.13.011034",
    journal = "Phys. Rev. X",
    volume = "13",
    number = "1",
    pages = "011034",
    year = "2023"
}

@article{Choi:2022jqy,
    author = "Choi, Yichul and Lam, Ho Tat and Shao, Shu-Heng",
    title = "{Noninvertible Global Symmetries in the Standard Model}",
    eprint = "2205.05086",
    archivePrefix = "arXiv",
    primaryClass = "hep-th",
    reportNumber = "YITP-SB-2022-21, MIT-CTP/5433",
    doi = "10.1103/PhysRevLett.129.161601",
    journal = "Phys. Rev. Lett.",
    volume = "129",
    number = "16",
    pages = "161601",
    year = "2022"
}

@article{Damia:2022bcd,
    author = "Damia, Jeremias Aguilera and Argurio, Riccardo and Garcia-Valdecasas, Eduardo",
    title = "{Non-Invertible Defects in 5d, Boundaries and Holography}",
    eprint = "2207.02831",
    archivePrefix = "arXiv",
    primaryClass = "hep-th",
    doi = "10.21468/SciPostPhys.14.4.067",
    journal = "SciPost Phys.",
    volume = "14",
    pages = "067",
    year = "2023"
}

@article{Cordova:2022rer,
    author = "Cordova, Clay and Ohmori, Kantaro and Rudelius, Tom",
    title = "{Generalized symmetry breaking scales and weak gravity conjectures}",
    eprint = "2202.05866",
    archivePrefix = "arXiv",
    primaryClass = "hep-th",
    doi = "10.1007/JHEP11(2022)154",
    journal = "JHEP",
    volume = "11",
    pages = "154",
    year = "2022"
}

@article{Apruzzi:2022dlm,
    author = "Apruzzi, Fabio",
    title = "{Higher form symmetries TFT in 6d}",
    eprint = "2203.10063",
    archivePrefix = "arXiv",
    primaryClass = "hep-th",
    doi = "10.1007/JHEP11(2022)050",
    journal = "JHEP",
    volume = "11",
    pages = "050",
    year = "2022"
}

@article{Apruzzi:2022rei,
    author = "Apruzzi, Fabio and Bah, Ibrahima and Bonetti, Federico and Schafer-Nameki, Sakura",
    title = "{Noninvertible Symmetries from Holography and Branes}",
    eprint = "2208.07373",
    archivePrefix = "arXiv",
    primaryClass = "hep-th",
    doi = "10.1103/PhysRevLett.130.121601",
    journal = "Phys. Rev. Lett.",
    volume = "130",
    number = "12",
    pages = "121601",
    year = "2023"
}

@article{GarciaEtxebarria:2022vzq,
    author = "Garc\'\i{}a Etxebarria, I\~naki",
    title = "{Branes and Non-Invertible Symmetries}",
    eprint = "2208.07508",
    archivePrefix = "arXiv",
    primaryClass = "hep-th",
    doi = "10.1002/prop.202200154",
    journal = "Fortsch. Phys.",
    volume = "70",
    number = "11",
    pages = "2200154",
    year = "2022"
}

@article{Heckman:2022muc,
    author = {Heckman, Jonathan J. and H\"ubner, Max and Torres, Ethan and Zhang, Hao Y.},
    title = "{The Branes Behind Generalized Symmetry Operators}",
    eprint = "2209.03343",
    archivePrefix = "arXiv",
    primaryClass = "hep-th",
    doi = "10.1002/prop.202200180",
    journal = "Fortsch. Phys.",
    volume = "71",
    number = "1",
    pages = "2200180",
    year = "2023"
}

@article{Kaidi:2022cpf,
    author = "Kaidi, Justin and Ohmori, Kantaro and Zheng, Yunqin",
    title = "{Symmetry TFTs for Non-Invertible Defects}",
    eprint = "2209.11062",
    archivePrefix = "arXiv",
    primaryClass = "hep-th",
    month = "9",
    year = "2022"
}

@article{Antinucci:2022vyk,
    author = "Antinucci, Andrea and Benini, Francesco and Copetti, Christian and Galati, Giovanni and Rizi, Giovanni",
    title = "{The holography of non-invertible self-duality symmetries}",
    eprint = "2210.09146",
    archivePrefix = "arXiv",
    primaryClass = "hep-th",
    reportNumber = "SISSA 16/2022/FISI",
    month = "10",
    year = "2022"
}

@article{Choi:2022fgx,
    author = "Choi, Yichul and Lam, Ho Tat and Shao, Shu-Heng",
    title = "{Non-invertible Gauss Law and Axions}",
    eprint = "2212.04499",
    archivePrefix = "arXiv",
    primaryClass = "hep-th",
    reportNumber = "MIT-CTP/5504, YITP-SB-2022-39",
    month = "12",
    year = "2022"
}

@article{Heckman:2022xgu,
    author = "Heckman, Jonathan J. and Hubner, Max and Torres, Ethan and Yu, Xingyang and Zhang, Hao Y.",
    title = "{Top Down Approach to Topological Duality Defects}",
    eprint = "2212.09743",
    archivePrefix = "arXiv",
    primaryClass = "hep-th",
    month = "12",
    year = "2022"
}

@article{Kaidi:2023maf,
    author = "Kaidi, Justin and Nardoni, Emily and Zafrir, Gabi and Zheng, Yunqin",
    title = "{Symmetry TFTs and Anomalies of Non-Invertible Symmetries}",
    eprint = "2301.07112",
    archivePrefix = "arXiv",
    primaryClass = "hep-th",
    month = "1",
    year = "2023"
}

@article{Etheredge:2023ler,
    author = "Etheredge, Muldrow and Garcia Etxebarria, Inaki and Heidenreich, Ben and Rauch, Sebastian",
    title = "{Branes and symmetries for $\mathcal N=3$ S-folds}",
    eprint = "2302.14068",
    archivePrefix = "arXiv",
    primaryClass = "hep-th",
    month = "2",
    year = "2023"
}

@article{Zhang:2023wlu,
    author = "Zhang, Carolyn and C\'ordova, Clay",
    title = "{Anomalies of $(1+1)D$ categorical symmetries}",
    eprint = "2304.01262",
    archivePrefix = "arXiv",
    primaryClass = "cond-mat.str-el",
    month = "4",
    year = "2023"
}

@article{Hsin:2020nts,
    author = "Hsin, Po-Shen and Lam, Ho Tat",
    title = "{Discrete theta angles, symmetries and anomalies}",
    eprint = "2007.05915",
    archivePrefix = "arXiv",
    primaryClass = "hep-th",
    doi = "10.21468/SciPostPhys.10.2.032",
    journal = "SciPost Phys.",
    volume = "10",
    number = "2",
    pages = "032",
    year = "2021"
}

@article{Cordova:2018cvg,
    author = "C\'ordova, Clay and Dumitrescu, Thomas T. and Intriligator, Kenneth",
    title = "{Exploring 2-Group Global Symmetries}",
    eprint = "1802.04790",
    archivePrefix = "arXiv",
    primaryClass = "hep-th",
    doi = "10.1007/JHEP02(2019)184",
    journal = "JHEP",
    volume = "02",
    pages = "184",
    year = "2019"
}

@article{Witten:1998wy,
    author = "Witten, Edward",
    title = "{AdS / CFT correspondence and topological field theory}",
    eprint = "hep-th/9812012",
    archivePrefix = "arXiv",
    reportNumber = "IASSNS-HEP-98-96",
    doi = "10.1088/1126-6708/1998/12/012",
    journal = "JHEP",
    volume = "12",
    pages = "012",
    year = "1998"
}

@article{Lee:2021crt,
    author = "Lee, Yasunori and Ohmori, Kantaro and Tachikawa, Yuji",
    title = "{Matching higher symmetries across Intriligator-Seiberg duality}",
    eprint = "2108.05369",
    archivePrefix = "arXiv",
    primaryClass = "hep-th",
    doi = "10.1007/JHEP10(2021)114",
    journal = "JHEP",
    volume = "10",
    pages = "114",
    year = "2021"
}

@article{Apruzzi:2021vcu,
    author = "Apruzzi, Fabio and Schafer-Nameki, Sakura and Bhardwaj, Lakshya and Oh, Jihwan",
    title = "{The Global Form of Flavor Symmetries and 2-Group Symmetries in 5d SCFTs}",
    eprint = "2105.08724",
    archivePrefix = "arXiv",
    primaryClass = "hep-th",
    doi = "10.21468/SciPostPhys.13.2.024",
    journal = "SciPost Phys.",
    volume = "13",
    number = "2",
    pages = "024",
    year = "2022"
}

@article{Apruzzi:2021mlh,
    author = "Apruzzi, Fabio and Bhardwaj, Lakshya and Gould, Dewi S. W. and Schafer-Nameki, Sakura",
    title = "{2-Group symmetries and their classification in 6d}",
    eprint = "2110.14647",
    archivePrefix = "arXiv",
    primaryClass = "hep-th",
    doi = "10.21468/SciPostPhys.12.3.098",
    journal = "SciPost Phys.",
    volume = "12",
    number = "3",
    pages = "098",
    year = "2022"
}

@article{Cvetic:2023plv,
    author = {Cveti\v{c}, Mirjam and Heckman, Jonathan J. and H\"ubner, Max and Torres, Ethan},
    title = "{Fluxbranes, Generalized Symmetries, and Verlinde's Metastable Monopole}",
    eprint = "2305.09665",
    archivePrefix = "arXiv",
    primaryClass = "hep-th",
    month = "5",
    year = "2023"
}

@article{DelZotto:2022joo,
    author = "Del Zotto, Michele and Garc\'\i{}a Etxebarria, I\~naki and Schafer-Nameki, Sakura",
    title = "{2-Group Symmetries and M-Theory}",
    eprint = "2203.10097",
    archivePrefix = "arXiv",
    primaryClass = "hep-th",
    doi = "10.21468/SciPostPhys.13.5.105",
    journal = "SciPost Phys.",
    volume = "13",
    pages = "105",
    year = "2022"
}

@article{Cvetic:2022imb,
    author = {Cveti\v{c}, Mirjam and Heckman, Jonathan J. and H\"ubner, Max and Torres, Ethan},
    title = "{0-form, 1-form, and 2-group symmetries via cutting and gluing of orbifolds}",
    eprint = "2203.10102",
    archivePrefix = "arXiv",
    primaryClass = "hep-th",
    reportNumber = "UPR-1317-T, CERN-TH-2022-053",
    doi = "10.1103/PhysRevD.106.106003",
    journal = "Phys. Rev. D",
    volume = "106",
    number = "10",
    pages = "106003",
    year = "2022"
}

@article{Genolini:2022mpi,
    author = "Genolini, Pietro Benetti and Tizzano, Luigi",
    title = "{Comments on Global Symmetries and Anomalies of $5d$ SCFTs}",
    eprint = "2201.02190",
    archivePrefix = "arXiv",
    primaryClass = "hep-th",
    month = "1",
    year = "2022"
}
\bibliographystyle{ytphys}

\end{document}